\documentclass[lettersize,journal]{IEEEtran}
\usepackage{amsmath,amsfonts}
\usepackage{algorithmic}
\usepackage{algorithm}
\usepackage{array}
\usepackage[caption=false,font=normalsize,labelfont=sf,textfont=sf]{subfig}
\usepackage{textcomp}
\usepackage{stfloats}
\usepackage{url}
\usepackage{verbatim}
\usepackage{graphicx}
\usepackage{cite}
\usepackage{soul}
\usepackage{amssymb}
\usepackage{amsmath}
\usepackage{cite}
\usepackage{balance}
\usepackage{booktabs}
\usepackage{float}
\usepackage{graphicx}
\usepackage{multirow}
\usepackage{footnote}
\usepackage{stfloats}
\usepackage{array}

\usepackage{amsmath} 
\usepackage{amsfonts} 
\usepackage{url}
\usepackage{balance}

\usepackage[misc]{ifsym} 
\usepackage{colortbl}  %彩色表格需要加载的宏包
\usepackage{xcolor}
\usepackage{array}   %对表列和表格线的设置需要用到array宏包

\newcommand{\rr}[1]{\textcolor{black}{#1}}
\newcommand{\hyss}[1]{\textcolor{black}{#1}}
\newcommand{\tcmi}[1]{\textcolor{black}{#1}}
\newcommand{\eg}{\textit{e.g.}}

\newcommand{\ie}{\textit{i.e.}}
\newcommand{\RNum}[1]{\uppercase\expandafter{\romannumeral #1\relax}}

\begin{document}

\title{Infrared Image Super-Resolution: A Systematic Review and Future Trends}

\author{Yongsong~Huang,~\IEEEmembership{Member,~IEEE,}
~Tomo~Miyazaki,~\IEEEmembership{Member,~IEEE,}%  
~Xiaofeng ~Liu,~\IEEEmembership{Member,~IEEE,}
        and ~Shinichiro~Omachi,~\IEEEmembership{Senior Member,~IEEE}%
        % <-this % stops a space
\thanks{This paper was produced by the IEEE Publication Technology Group. They are in Piscataway, NJ.}% <-this % stops a space
\thanks{Manuscript received April 19, 2021; revised August 16, 2021.}}

\markboth{Journal of \LaTeX\ Class Files,~Vol.~14, No.~8, August~2021}%
{Shell \MakeLowercase{\textit{et al.}}: A Sample Article Using IEEEtran.cls for IEEE Journals}


\maketitle

% \begin{abstract}
% Image Super-Resolution (SR) is essential for a wide range of computer vision and image processing tasks. Tackling the challenge of infrared (IR) image (or thermal image) super-resolution is a key and persistent research area within the field of deep learning. This survey aims to provide a comprehensive perspective of IR image super-resolution, including its applications, hardware imaging system dilemmas, and taxonomy of image processing methodologies. In addition, the datasets and evaluation metrics in IR image super-resolution tasks are also discussed. Furthermore, the deficiencies in current technologies and possible promising directions for the community to explore are highlighted. To cope with the rapid development in this field, we intend to regularly update the relevant excellent work at \url{https://github.com/yongsongH/Infrared_Image_SR_Survey}.
% \end{abstract}

\begin{abstract}
Image Super-Resolution (SR) is a fundamental task in computer vision and image processing. The challenge of super-resolving infrared (IR) or thermal images represents a key and persistent research frontier in the deep learning era. \rr{Different from visible-light SR, IRSR is challenged by the distinct properties of thermal images, which typically exhibit low contrast, limited high-frequency details, and sensor-specific noise.} This survey provides a comprehensive review of Infrared Image Super-Resolution (IRSR), covering its applications, the inherent limitations of IR imaging hardware, and a detailed taxonomy of processing methodologies. In addition, we discuss the critical roles of datasets and evaluation metrics in the field. Finally, we highlight current technological gaps and identify promising future directions for the community. To keep pace with the rapid developments, we maintain an actively updated repository of relevant work at \url{https://github.com/yongsongH/Infrared_Image_SR_Survey}.
\end{abstract}

\begin{IEEEkeywords}
Image Super-resolution, Deep Learning, Convolutional Neural Networks (CNN), Generative Adversarial Nets 
\end{IEEEkeywords}

\section{Introduction}\label{sec:introduction}

\IEEEPARstart{I}{mage} super-resolution (SR) is a classic low-level vision task in image processing\cite{wang2020deep,park2003super,chen2022real,liu2022blind}. The super-resolution reconstruction of IR images aims to recover high-resolution (HR) IR images from low-resolution (LR) IR images\cite{wang2020deep,park2003super}. Due to the existence of diverse degradation paths, the super-resolution reconstruction for IR images is still considered to be an ill-posed problem\cite{yang2008image,yang2010image,park2003super}. More details of the problem definition are shown in Eq. \ref{eq.1}:

% \begin{equation}
% I_{LR}=\mathbb{D}(I_{HR};\delta )
% \label{eq.1}
% \end{equation} where $\mathbb{D}$ denotes a degradation function, $I_{HR}$ is the high-resolution IR image, $I_{LR}$ is the low-resolution IR image and $\delta$ is the parameters of the degradation process. In general, the degradation process includes the following factors: 1) downsampling, 2) noise, and 3) compression, as shown in Eq. \ref{eq.2}. In different real-world settings, the importance of these factors will be different.

% \begin{equation}
% \mathbb{D}\left(I_{HR} ; \delta\right)=\left(I_{HR} \otimes \kappa\right) \downarrow_d+n_{\varsigma},\{\kappa, s, \varsigma\} \subset \delta
% \label{eq.2}
% \end{equation} where $I_{HR} \otimes {k} $ represents the convolution between a blur kernel ${k}$ and the $\mathrm{HR}$ image $I_{HR}$. In the ${k}$, noise and compression are included. And, $\downarrow _{d}$  is a downsampling factor, \eg, $4 \times$ and $8 \times$. $n_{\varsigma}$ is some additive white Gaussian noise with standard deviation $\varsigma$. Briefly, the super-resolution reconstruction objective function of IR images can be described as:

\begin{equation}
I_{LR}=\mathbb{D}(I_{HR};\delta )
\label{eq.1}
\end{equation} \rr{where $I_{HR}$ is the latent high-resolution IR image, $I_{LR}$ is its observed low-resolution counterpart, and $\delta$ represents the parameters of the degradation process. This general model is commonly specified with a formulation that accounts for blurring, downsampling, and noise, as shown in Eq. \ref{eq.2}.} 
\begin{equation}
\mathbb{D}\left(I_{HR} ; \delta\right)=\left(I_{HR} \otimes \kappa\right) \downarrow_d+n_{\varsigma},\{\kappa, d, \varsigma\} \subset \delta
\label{eq.2}
\end{equation} 
\rr{Here, each component models a specific aspect of the image acquisition process. The term $\left(I_{HR} \otimes \kappa\right)$ represents a convolution operation between the HR image and a blur kernel $\kappa$. This kernel models the point spread function of the imaging system, which originates from physical factors like lens imperfections (optical blur), atmospheric turbulence, or motion blur. In non-blind SR settings, $\kappa$ is often assumed to be a known, simple function, such as an isotropic Gaussian kernel\cite{wang2020deep}. In contrast, blind SR approaches treat $\kappa$ as unknown and attempt to estimate it from the LR image\cite{park2003super,chen2022real}. The operator $\downarrow_d$ is a downsampling operation with a scale factor $d$. This models the limited spatial resolution of the imaging sensor, which samples a continuous scene onto a discrete pixel grid. In the SR task, this process is most commonly simulated using bicubic interpolation, an assumption that underlies the creation of many standard benchmark datasets. The term $n_{\varsigma}$ represents additive noise introduced during acquisition, primarily from sensor electronics (e.g., thermal and readout noise). It is typically assumed to be Additive White Gaussian Noise with a standard deviation of $\varsigma$\cite{yang2008image,yang2010image}. The super-resolution reconstruction objective function of IR images can be described as:}

% where $I_{HR} \otimes {k} $ represents the convolution between a blur kernel ${k}$ and the $\mathrm{HR}$ image $I_{HR}$. In the ${k}$, noise and compression are included. And, $\downarrow _{d}$  is a downsampling factor, \eg, $4 \times$ and $8 \times$. $n_{\varsigma}$ is some additive white Gaussian noise with standard deviation $\varsigma$. Briefly, the super-resolution reconstruction objective function of IR images can be described as:

\begin{equation}
\hat{\theta}=\underset{\theta}{\arg \min } \mathcal{L}\left(I_{HR}, I_{SR}\right)+\lambda \Phi(\theta)
\label{eq.3}
\end{equation} where $\mathcal{L}$ denotes the loss function, between the HR image $I_{HR}$ and the SR image $I_{SR}$. $\Phi(\theta)$ and $\lambda$ are the regularization term and punishment parameter, respectively. Initial works follow the IR images SR as general image SR, and simply adopt the visible images SR model (\eg, Convolutional Neural Networks (CNN)-based methods\cite{dong2015image}, Generative Adversarial Nets (GAN)-based methods\cite{ledig2017photo,wang2020deep,wang2021real}, and transformer model\cite{liang2021swinir}) to the new task. 

\begin{figure*}[!t]
\centering
\includegraphics[width=\textwidth]{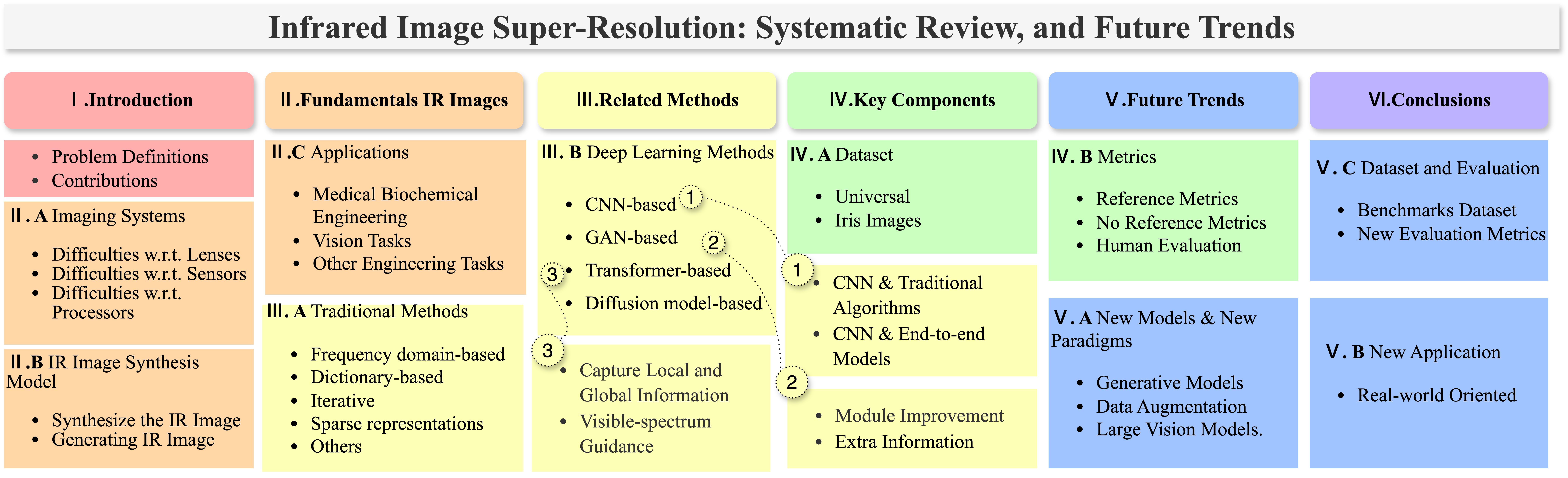}%
\label{fig_first_case}
\caption{A hierarchical taxonomy of the Infrared Image Super-Resolution field. This figure provides a structured overview of the key research areas, methodologies, and challenges discussed in this survey, serving as a visual guide to the paper's content and the domain's landscape.}
\label{fig3}
\end{figure*}

However, it should be pointed out that the degradation process in infrared images is more complicated. The main reasons include the following: first, the blur kernel used in the reconstruction of visible images is relatively simple. It is typically Gaussian noise and JPEG compression\cite{gao2022maritime,wang2021real,luo2022learning,zhang2021designing,prajapati2021channel}. But, these blur kernels are not robust enough for IR images to represent complex real-world conditions. Because IR imaging systems are limited by optical physics and electronics. In other words, the noise generated is different. In addition, many IR cameras are deployed in the wild environment. It further increases the complexity of IR image degradation. On the other hand, the IR image presents simple color and insignificant gradients. A fundamental challenge is the significant "information overlap between high and low frequencies"\cite{gao2022maritime,wang2022compressed,wang2021super,yang2020deep,almasri2018multimodal,suryanarayana2019infrared,gungor2019transform,zhang2019super,liu2019infrared,zhang2018infrared,gang2013study,han2018infrared,marivani2020multimodal}. In visible-light images, high-frequency components (e.g., sharp edges, fine textures) are typically distinct from low-frequency components (e.g., smooth areas). In contrast, due to sensor limitations and thermal diffusion, IR images are inherently blurrier. This causes the weak, high-frequency details to become spectrally mixed with the dominant low-frequency background and noise. Consequently, it is much harder for an SR model to distinguish genuine fine structures from noise artifacts, a problem that is less severe in the visible spectrum. These characteristics, which are different from visible images, can be considered to be unique patterns in IR images (see Sec.\ref{sec3}). Consequently, it is necessary to propose super-resolution reconstruction approaches appropriate for IR images. We will discuss more details in future sections.

% In this paper, we systematically review applications, methods, and challenges in the field of IR image super-resolution. We aim to conduct a comprehensive review and classification of IR image super-resolution methods. Various types of IR images, ranging from short-wave to long-wave IR images, are considered and discussed in terms of the various super-resolution techniques employed. 

\tcmi{In this paper, we systematically review the applications, methods, and challenges in IR image super-resolution, targeting both academic researchers and industrial practitioners. This survey aims to bridge the gap between traditional SR methods and the unique characteristics of IR imaging, offering actionable insights for developing domain-specific algorithms. For academic researchers, it provides an in-depth analysis of innovative methodologies and open challenges. For industrial practitioners, it highlights practical challenges, such as hardware limitations and deployment in dynamic environments, with guidance for optimizing IR imaging applications in fields like autonomous driving, medical imaging, and remote sensing. By systematically analyzing existing approaches and identifying gaps, this survey serves as a foundational reference to advance research and practice in IR image super-resolution.} Unlike other reviews, our survey focuses on recent deep-learning algorithms and summarizes the specific patterns used in IR images for model representation. It will benefit the whole community to propose algorithms that are different from visible image reconstruction to improve the quality of IR images. To the best of our knowledge, this is also the first work that provides the most detailed explanation of why enhancing hardware systems to improve IR image quality is difficult. Unlike existing SR surveys\cite{wang2020deep,park2003super,chen2022real,wang2021deepl,danaci2022survey}, this survey uses the perspective of the IR image's unique patterns to illustrate the challenges that distinguish it from visible images and also aims to review recent advances in SR techniques for IR images in the deep learning field. Meanwhile, we also summarized the application of IR images in various engineering fields. It will help the researchers to have a comprehensive insight into the entire field. The hierarchical and structural taxonomy of this survey is shown in Fig.\ref{fig3}. \rr{The hierarchical and structural taxonomy of this survey is shown in Fig.\ref{fig3}. This structure is designed to guide the reader logically through the domain, from foundational concepts to forward-looking challenges. We begin by establishing the Fundamentals of IR imaging and its applications, which provide the necessary context for understanding the Related Methods developed to solve the SR problem. Subsequently, we detail the essential tools for empirical validation—Datasets and Metrics—before concluding with a discussion on the unique Challenges and Future Trends that will shape the field.}

\begin{figure*}[!t]
\centering
\includegraphics[width=\textwidth]{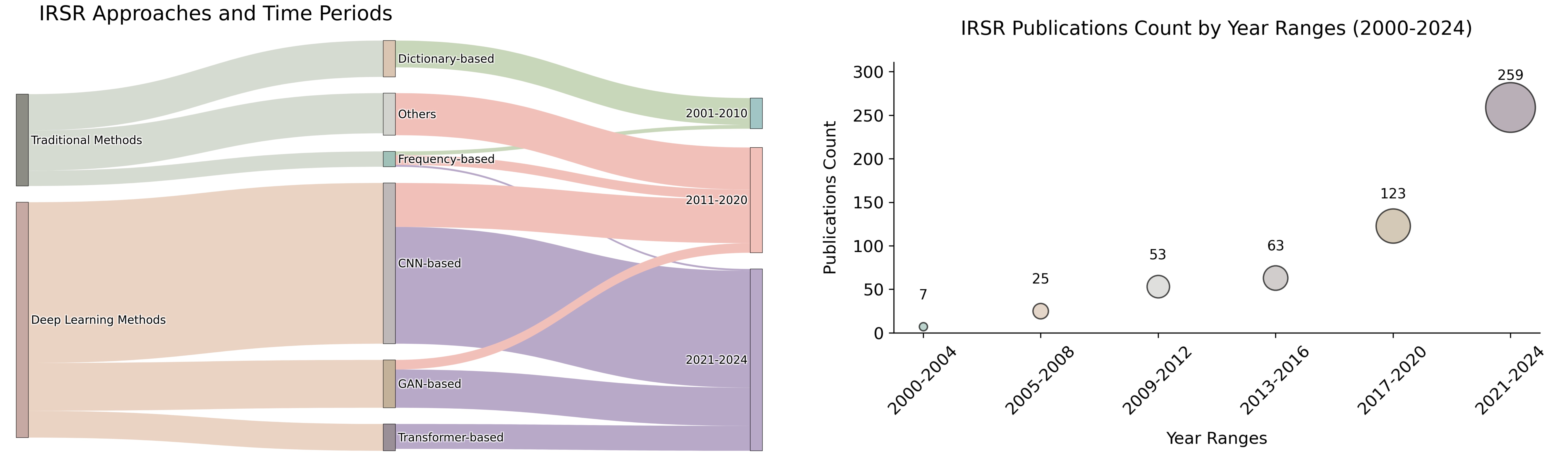}%
\label{fig_first_case}
\caption{\tcmi{[left sub-figure] Transition of IRSR methodologies across periods; [right sub-figure] Number of papers released to date. This analysis includes publications indexed in the Web of Science Core Collection from 2000 to 2024.}}
\label{fig_time}
\end{figure*}

\tcmi{The left sub-figure in Fig.\ref{fig_time} illustrates the evolution of methodologies for IRSR, categorizing them into traditional approaches, such as dictionary-based and frequency-based methods, and more recent deep learning-based techniques. Traditional methods dominated from 2000 to 2010, focusing primarily on frequency-domain modeling and sparse representation. Starting from 2011, the introduction of CNNs brought significant advancements, establishing their dominance for nearly a decade. More recently, GANs have emerged as a powerful tool for reconstructing high-quality textures, particularly excelling in applications requiring realistic detail synthesis, such as enhancing edges and textures in low-contrast IR images. However, their sensitivity to domain shifts often results in artifacts. In contrast, Transformer-based approaches\cite{mehri2023tntvit,shi2024swinibsr} excel in global contextual modeling, making them more robust for tasks requiring structural consistency and long-range dependency modeling, such as IR image reconstruction in dynamic environments. This progression highlights the continuous innovation in IRSR methodologies to address the unique challenges posed by IR images. Compared to traditional frequency-domain methods, which focus on enhancing high-frequency details, deep learning-based methods such as CNNs\cite{zhang2024dual} and GANs\cite{han2023dual} excel in learning non-linear mappings and synthesizing realistic textures. However, traditional methods remain advantageous in resource-constrained scenarios due to their lower computational complexity. Meanwhile, Transformer-based\cite{mehri2023tntvit} and state-space models\cite{huang2024irsrmamba} bring further advancements by enabling global contextual modeling and efficient scalability, making them particularly suitable for applications involving complex noise patterns and high-resolution requirements.}

\tcmi{The right sub-figure presents the publication trends in the field of IRSR from 2000 to 2024. It demonstrates an exponential growth in research output, particularly from 2017 to 2024, where the number of publications rose sharply from 123 to 259. This surge reflects not only the technological advancements achieved through the adoption of deep learning-based methods, such as GANs and Transformers, but also the expanding practical applications of IR images in areas like autonomous driving, medical imaging, and remote sensing. These trends underscore the rapid evolution of IRSR research, making this survey a timely resource for researchers and practitioners aiming to navigate this dynamic field.}

This survey's contribution can be summarized as the following:

\textbf{1)} We systematically review the applications, methods, datasets, and evaluation metrics related to IR image super-resolution. In particular, we carefully investigated the application in IR image super-resolution using relevant methods since the beginning of the deep learning explosion. It includes the classification and summarization of these methods. \textbf{2)} We also provide some unique pattern details in IR images that will benefit reconstruction, which are features making IR images super-resolution different from visible image super-resolution. It is common to see poor performance when directly transferring methods from visible images to IR images. It is promising to be solved by focusing more on the unique features of IR images. \textbf{3)} We discuss trends, open issues, and challenges in IR image super-resolution. This section is dedicated to providing the community with new perspectives and new directions to explore. We will continue to update our open-source repository with new work in this area, and hope it will be useful for future highlighting research. The repository link is \url{https://github.com/yongsongH/Infrared_Image_SR_Survey}. To ensure comprehensive coverage, this narrative review was compiled based on a systematic literature search across major academic databases, including IEEE Xplore, ACM Digital Library, SpringerLink, and Google Scholar, spanning from approximately 2000 to early 2024. Search keywords included "infrared image super-resolution," "thermal image super-resolution," "IRSR," "deep learning," "generative adversarial network," and "transformer" in combination with "infrared" or "thermal." While not adhering to a strict PRISMA protocol, this process was designed to capture seminal traditional works as well as the full spectrum of recent deep learning advancements.

\tcmi{The remainder of this survey is organized as follows: Section 2 introduces the fundamentals of IR images and IR image applications; Section 3 describes image processing methods for IR image super-resolution in the literature; Section 4 presents the key components in IR image super-resolution: datasets and metrics; Section 5 discusses IR image specific patterns, future trends, and open issues; and Section 6 summarizes the work.}

\section{Fundamentals of Infrared Images}

In this section, we will introduce the fundamentals of IR images, which are IR imaging systems and applications. For the imaging system, we will discuss the components in the system and how these items influence the IR image quality. \hyss{Then, the physics model of the infrared imaging quality limitation will be presented.} Finally, IR image applications in different engineering fields will be classified and then discussed. 

\begin{figure}[t]
\centerline{\includegraphics[width=\columnwidth]{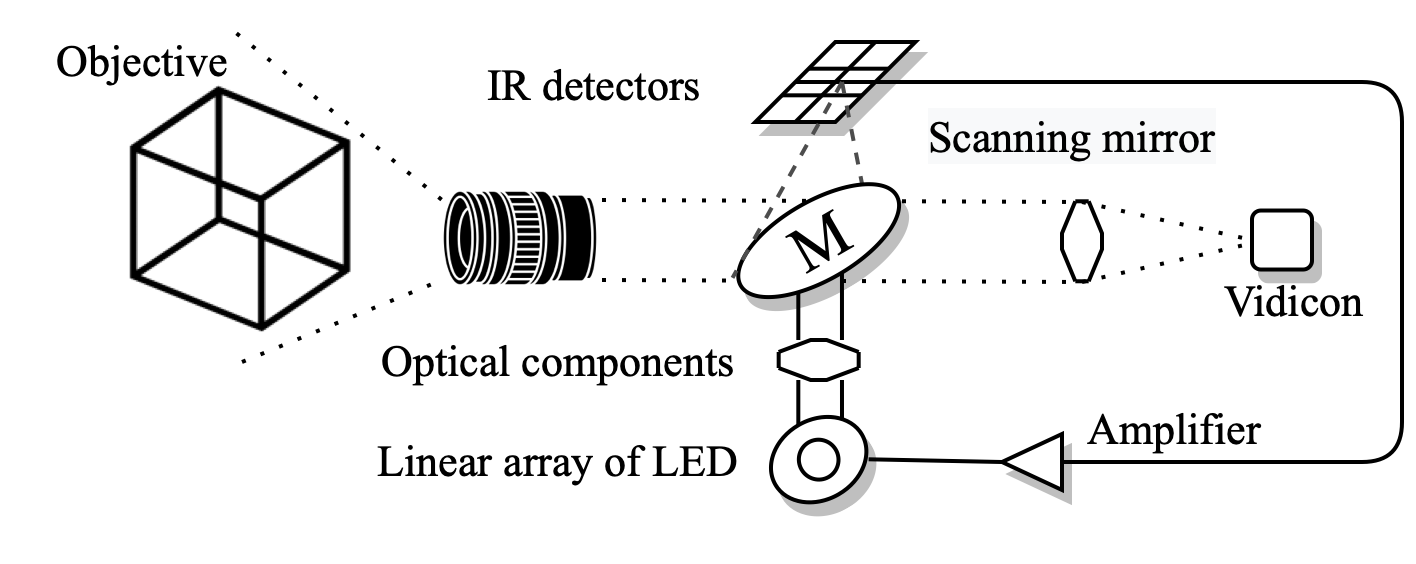}}
%0.8\textwidth
\caption{Explanation of the operation mechanism in a FLIR system.}
\label{fig2}
\end{figure}
\vspace{-0.5cm}

\subsection{Imaging Systems\label{sec.2.1}}

In the classical IR imaging system, the following basic components are included: lenses, sensors, processors, and others\cite{park2003super,laikin2018lens}. In Fig. \ref{fig2}, we show the details in a classic forward-looking infrared (FLIR) system. On the linear array of IR detectors, the objective produces an image. Each detector's amplified output drives a corresponding light-emitting diode (LED). After reflection from the back of the scanning mirror, this LED array is transferred onto an image tube (vidicon). In turn, this powers a typical cathode ray tube display. The whole system converts a scanned infrared image into a digital representation that can be seen on a monitor\cite{laikin2018lens}. At present, the most popular infrared imaging systems use incorporate solid-state detectors.

We can achieve the goal, which is improving the IR image quality, by enhancing the performance of imaging systems. According to current research, some key components in the imaging system: \textbf{lenses, sensors, and processors,} are significant in improving IR image quality. However, we will face many challenges in improving the image quality directly from the imaging system. More details will be discussed in the next sections.

\subsubsection{Difficulties w.r.t. Lenses}

In general, we need more light to be captured by the imaging system to get better image quality\cite{laikin2018lens,yuan2018compressive}. But more light means a larger lens, which brings two main challenges: First, the more sophisticated instrumentation required will increase the costs and difficulties in the entire industrial design\cite{rabal2018dynamic,lee2007image}. Secondly, larger optical lenses make the size of the detector also larger\cite{yuan2018compressive}. It makes the whole size of the imaging system becomes larger and the combination of both becomes more difficult\cite{mas2020study}. In motion situations, such as fast-moving environments, less light will arrive at the machine, making the image quality poor\cite{yuan2018compressive,bijl2012effectiveness}. \rr{These optical constraints directly contribute to a lower signal-to-noise ratio (SNR) and limited spatial resolution in the captured raw images, establishing a challenging baseline for any subsequent SR task.}

\subsubsection{Difficulties w.r.t. Sensors}

With the captured IR light, the sensor is a key component to translate the light-wave signal to quantified signal. This component also involves two following major difficulties in improving the imaging performance. Firstly, there is the inherent noise of the sensor\cite{kim2018averaging,wang2016stripe,chen2014super,kong2013near,schutte2003signal}. It is a noise called fixed pattern noise (FPN), which exists in the differences between bias voltages in column readout circuits of uncooked infrared sensors resulting in strong stripe noise which changes slowly in time\cite{wang2016stripe,ming2013effect}. \rr{This type of thermal noise is distinct from the random noise typically found in visible-light sensors and introduces structured artifacts that are particularly difficult for SR models to handle.} If we address this problem from hardware, it is almost impossible to solve it completely due to the physical limits\cite{sun2013pixel,kong2013near,kopeika2012sub}. Secondly, the limitation of device size makes it difficult for the current technology to make advances in the short term\cite{yuan2018compressive,chen2014super}.

\rr{These hardware-level imperfections pose a significant challenge for the accuracy of image synthesis models used in infrared imaging. The complex nature of sensor-level issues, such as thermal noise and non-uniformity across the focal plane array, is notoriously difficult to model accurately. Consequently, synthetic IR data generation often relies on simplified noise models (e.g., additive Gaussian noise) that fail to capture the spatially correlated, non-linear, and often temperature-dependent characteristics of real-world sensor noise. This discrepancy between simplified synthetic data and the complex reality of captured IR images creates a significant simulation-to-reality domain gap. Models trained on such synthetic data may struggle to generalize to real-world scenarios, as they are not optimized to handle the unique artifact patterns originating from focal plane array non-uniformity and other sensor-specific defects, thereby limiting their practical accuracy.}

\subsubsection{Difficulties w.r.t. Processors}

For many IR imaging systems, the traditional microprocessors are still equipped. Therefore, it may be difficult to run the sophisticated SR algorithms at the stage of image capturing, because of the limited computing power itself\cite{young2006superresolution,sun2013pixel}.

% Improving the performance of the equipment can be promptly expressed in the image quality improvement. However, as mentioned above, we will face many difficulties with the key components used in IR camera imaging systems. It mainly includes the challenge of the fabrication process and physical limitations. For the fabrication process, we expect the camera to be more portable but often require larger lenses for better imaging. For processors, we need new technologies to improve computing and processing power, which requires more time and effort. 

\rr{Improving the performance of the equipment can be promptly expressed in the image quality improvement. However, as mentioned above, we will face many difficulties with the key components used in IR camera imaging systems, which include challenges in the fabrication process and inherent physical limitations. For instance, the demand for portability conflicts with the need for larger lenses for better imaging, and advancing processor technology requires significant time and effort. These hardware constraints inevitably lead to the captured images containing various forms of sensor noise and artifacts. Therefore, a critical challenge is to enhance the robustness of the super-resolution model itself. Future efforts must focus on improving the model's resilience to such real-world degradations, for example, through targeted data augmentation that simulates sensor noise and by developing robust training objectives less sensitive to input corruptions.}

In summary, improving imaging quality from hardware will be costly in terms of industrial expense and effort, and the final performance improvement will be limited due to insurmountable physical obstacles. \hyss{Next, another important task in IR imaging: synthetic models, will also be discussed.}

\subsection{IR Image Synthesis Model\label{sec3}}

\hyss{
In this section, we will discuss the IR image synthesis model in IR images. First, we will review previous work on synthesizing IR images. Then, the challenges of low-quality IR images that differ from normal images will be described. Finally, the connection between the IR image's unique patterns and the synthetic model will be presented.}

\subsubsection{Synthesize the IR Image}

\hyss{IR image quality relies on the proposed coloring physical model, which differs significantly from the normal image imaging in the visible wavelength band\cite{yu1998infrared}. Briefly, the IR image color first requires the sensor to capture the radiance from the object, and then depends on the proposed coloring model. However, the IR image quality will be limited because of the heat transfer model or the coloring method is incomplete. Compared with normal images, which can be imaged by sensor capture with visible light, the influence from the heat transfer model or coloring approach is removed. In the following, the IR image synthesis method will be discussed.}

\textit{Chanhee Oh's} work pioneered the use of octree for precise IR imaging of objects, but it overlooked non-homogeneous materials and internal heat sources\cite{oh1989integrated}. \textit{Nandhakumar's} method addressed this by simulating internal heat source effects using a hierarchical Vs-tree model, but it struggled with complex shapes due to extensive node computations\cite{nandhakumar1994unified}. \textit{Hyum-Ki Hong's} model represented heat transfer in non-homogeneous objects as an equivalent thermal circuit, effectively demonstrating the influence of the internal heat source\cite{hong1996simulation}. However, it calculated surface radiance by only considering convection and internal heat source conduction, neglecting heat flow effects between adjacent facets.

\rr{Recent research in physics-based infrared modeling has advanced beyond foundational principles to address more complex and specific phenomena through sophisticated computational methods. For instance, \textit{Li's} work developed a real-time infrared imaging simulation that accurately models the physical effects of sensors in both spatial and frequency domains, leveraging GPU-based parallel processing to achieve high fidelity without sacrificing real-time performance\cite{li2016real}. In the area of material properties, \textit{Xiong's} work employed a novel photon-tracking technique within a geometric optics framework to model the thermal infrared emissivity of snow and ice. Their work provided a unified physical explanation for the material's sensitivity to density over grain size, resolving previous experimental discrepancies\cite{xiong2024modeling}. Furthermore, research has expanded to include complex environmental interactions, such as the work by \textit{Hu}, who integrated discrete element method with augmented reality infrared modeling to quantify the impact of anthropogenic surface changes on thermal radiation signatures\cite{hu2025influence}. These studies collectively illustrate a trend toward integrating advanced simulation techniques to build more precise and comprehensive physical models of infrared phenomena, from sensor characteristics to nuanced environmental interactions.}

\hyss{It is clear that an effective IR image synthesis model must include all heat transfer modes inside the object, as well as the dynamic interactions between the object surface and its surroundings. With the help of these models, our goal is to capture the surface temperature variance to further generate IR images.}

\subsubsection{Generating IR Image}

\hyss{To generate realistic IR images, the thermal radiance of an object's surface must first be calculated. This process is fundamentally governed by the principles of atmospheric physics and blackbody radiation\cite{yu1998infrared}.}

\rr{The core principle is that an object's thermal emission can be modeled using \textit{Planck's Law}, which relates the spectral radiance of a blackbody to its absolute temperature $T$. For practical IR imaging, this calculation is performed within specific atmospheric windows (e.g., 3-5 $\mu m$ and 8-14 $\mu m$), which are spectral bands where atmospheric attenuation is minimal. To make the computation tractable, the integral form of Planck's Law is often approximated. The resulting equation provides a direct method to calculate the spectral radiant emittance, $E$, for a given temperature and wavelength band:}

\vspace{-0.2cm}
\begin{equation}
\scalebox{0.8}{% 缩小比例为 0.8
$E \approx \frac{C_1}{C_2 / T} e^{-C_2 X / T}\left\{X^3+\frac{3}{C_2 / T}\left[X^2+\frac{2}{C_2 / T}\left(X+\frac{1}{C_2 / T}\right)\right]\right\}_{X=\frac{1}{\lambda_1}}^{X=\frac{1}{\lambda_2}}$
}
\label{eq_rad}
\end{equation}

\rr{This approximation is highly accurate (relative error $<$ 1\%) under the conditions specified by \textit{Wien's Displacement Law} ($\lambda T<3000 \mu m \cdot K$), which are typically met in terrestrial IR imaging scenarios. The full derivation of Eq. \ref{eq_rad} from Planck's Law is provided in Appendix for interested readers. Once the radiance of each surface patch is determined using this model, these values are interpolated to the vertices and rendered, often using methods like Gouraud Shading, to produce the final synthetic IR image.}

To summarize, the enhancement of IR images faces significant challenges, primarily due to hardware constraints arising from physical limitations and the complexity of image synthesis models. Further, we can summarize the unique patterns in IR images from the model used to generated these images:

\begin{itemize}
    \item \textbf{Blur kernel.} The blur kernel employed in the reconstruction of visible images is comparatively straightforward, typically encompassing Gaussian noise and JPEG compression \cite{gao2022maritime,wang2021real,luo2022learning,zhang2021designing,prajapati2021channel}. However, these blur kernels lack the robustness required for IR images to accurately depict complex real-world scenarios or heat transfer model.
    \item \textbf{Degradation.} The deployment of IR cameras in uncontrolled, natural environments further exacerbates the complexity of IR image degradation.
    \item \textbf{Space information limited.} Because of the coloring model in IR image generating, the IR image presents simple color, insignificant gradients, and information overlap between high and low frequencies\cite{gao2022maritime,wang2022compressed,wang2021super,yang2020deep,almasri2018multimodal,suryanarayana2019infrared,gungor2019transform,zhang2019super,liu2019infrared,zhang2018infrared,gang2013study,han2018infrared,marivani2020multimodal}. Increased focus on these unique patterns will benefit to models applicable to IR images being proposed.
\end{itemize}

\begin{figure}[!t]
\centerline{\includegraphics[width=\columnwidth]{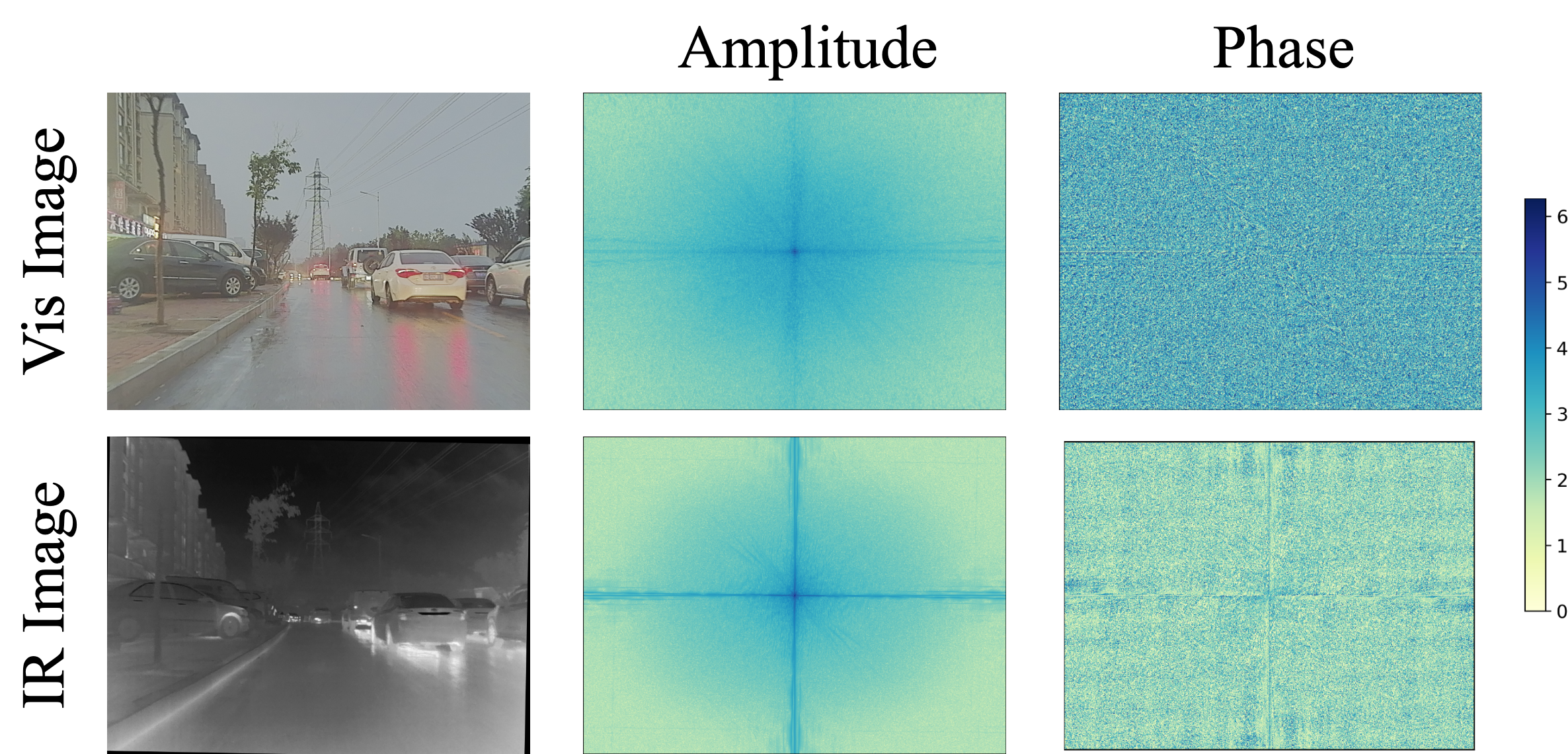}}
\caption{Comparison of Frequency Components in Visible and Infrared Images.}
\label{fig_4}
\end{figure}

\rr{To visually substantiate this concept, we present a frequency-domain analysis of a paired visible and infrared image in Fig.\ref{fig_4} (From M3FD Dataset\cite{Liu_2022_CVPR}). The Fourier transform decomposes an image into its amplitude and phase components, where the amplitude reflects style and low-level features, and the phase represents semantic structure. As shown in Fig.\ref{fig_4}, the amplitude spectrum of the visible image contains a wide distribution of energy extending from the low-frequency center outwards. This signifies a healthy presence of mid- and high-frequency information corresponding to the detailed textures of the road and the sharp edges of the vehicles. Conversely, the amplitude spectrum of the IR image shows energy that is intensely concentrated at the low-frequency center, with a rapid fall-off towards the higher frequencies. This spectral disparity provides clear evidence of the attenuated high-frequency content in IR images and underscores the fundamental challenge in IR super-resolution: recovering the rich details present in the visible domain from the low-frequency-dominant IR signal. Increased focus on these unique patterns will benefit to models applicable to IR images being proposed.}

\hyss{Based on the above concerns, we would resort to the subsequent image processing technology to improve the IR image quality. In the following sections, we will present more economical IRSR methods suitable for applications in various fields.}

\begin{table*}[t]
\centering
\caption{IR image super-resolution applications.}
\label{tab.1}
\renewcommand\arraystretch{1.5}
\resizebox{\textwidth}{!}{%
\begin{tabular}{@{}c|c|cc@{}}
\toprule
\textbf{Medical biochemical engineering} & \textbf{Vision tasks} & \multicolumn{2}{c}{\textbf{Other engineering tasks}} \\ \midrule
Pharmaceutical industry\cite{Offroy2012IncreasingTS,Offroy2010InfraredCI} & Image conversion\cite{Lee2022SuperresolutionTG}       & \multicolumn{1}{c|}{Automated vehicle\cite{Gupta2022TowardUG,shen2019depth,rukkanchanunt2017image}} & Food quality control\cite{martinez2022advantage,wang2021identification}      \\
Medical science\cite{Torra2022VersatileNS,CanalesFiscal2021COVID19CU,Lukose2021OpticalTF,liu2022infrared}         & Multispectral matching\cite{Bodensteiner2018MultispectralMU,Almasri2018MultimodalSF} & \multicolumn{1}{c|}{Remote sensing\cite{zhu2022pkulast,ouyang2022blockchain,luo2025learning,gong2025hsroadnet,song2023aednet}}    & Agricultural management\cite{zhu2022pkulast,geng2025evaluation,liu2022mapping,pang2024masked,yao2020water,wen2020segmentation,kerkech2020vine}   \\
Cellular observations\cite{Wang2018LiveCV,Benard2021OptimizationOA,dufour2006two}   & Targets detection\cite{neris2022fpga,zhao2022target,zhang2024coherent}      & \multicolumn{1}{c|}{Terrain models\cite{lin2021novel}}    & Water resource management\cite{lloyd2021optically,ping2021can,yang2020water} \\
Fluorescence microscopy\cite{lukose2021optical,mazaheri2022investigating,zhang2021deep,pavlovetc2020quantitative,li2015super} & Face recognition\cite{szankin2019influence,shao2009super,zhang2014face,zhang2015edge,redlich2016embedded}       & \multicolumn{1}{c|}{Land surface\cite{chen2021geometry,allred2021time,yaacob2020rock}}      & Star formation\cite{yamaguchi2020super,harvey2012first,biller2005high,megeath1996evidence}            \\ \bottomrule
\end{tabular}
}
\end{table*}
\vspace{-0.5cm}

\subsection{Applications}

IR image super-resolution reconstruction techniques have critical applications in various fields. In this section, we will show the applications in these key areas as shown in Tab. \ref{tab.1}. In summary, they can be generally divided into three categories: medical biochemical engineering, vision tasks, and other engineering tasks.

\subsubsection{Medical Biochemical Engineering}

For medical biochemical engineering, there are two typical examples presented here: neurodegenerative diseases and COVID-19 diagnosis. Super-resolution reconstructed infrared images can be used in the diagnosis of neurodegenerative diseases\cite{Torra2022VersatileNS}. As a non-invasive examination approach, it helps to visualize and provide evidence for consultation. The current global epidemic COVID-19 diagnosis also requires the help of IR images. Because CT images are costly for large-scale population screening, the use of infrared imaging can provide a replacement. The super-resolution reconstructed IR images are also used to extract time-series features, which are finally used as a reference for classification using machine learning methods\cite{Lukose2021OpticalTF,CanalesFiscal2021COVID19CU}. Moreover, high-resolution IR images are fundamental for the visualization of living cells\cite{Morozumi2020SpontaneouslyBF,Wang2018LiveCV}, protein structure analysis\cite{Barnett2017RepurposingAP,Baddeley20114DSM}, and nanoscale drug\cite{Proetto2018TumorRO}. These studies can be further developed with the help of super-resolution reconstruction in IR images.

\subsubsection{Vision Tasks} 
For vision tasks, super-resolution reconstruction of infrared images also plays an important role. We will present several representative tasks: multispectral matching, image conversion, and object detection. Multispectral matching\cite{Bodensteiner2018MultispectralMU,Almasri2018MultimodalSF} is popularly used in disaster prevention and homeland security. Better resolution of infrared images means better matching results. It is also beneficial for the object detection and other tasks. By using IR images, it is able to identify objects unaffected by the weather, illumination, or the environment's color. And, image conversion can convert electro-optical (EO) images to IR images\cite{Lee2022SuperresolutionTG}. The image conversion is significant because we use IR images for recognition, but we often receive EO images. On the other hand, IR images are very critical for object detection in some environments. For example, penetrating IR images are the key information source after disasters such as earthquakes. With the help of HR images, we can locate trapped people in dark environments, such as mines, by using localization algorithms\cite{Zhou2022YOLOSASEAI,Li2022ImprovedYN,Ramos2015EmbeddedSF}.

\subsubsection{Other Engineering Tasks}
For other engineering tasks, we can find the value of high-quality IR images in different disciplines. In the automated vehicle, infrared images in low-visibility conditions would be beneficial for assisted driving systems, but high-resolution thermal imaging sensors are typically expensive, which limits the general availability of such imaging systems\cite{Gupta2022TowardUG}. Therefore, super-resolution techniques can be used to improve performance in this filed\cite{Shen2019ADE, Rukkanchanunt2017ImageEF}. Further, super-resolution IR images have the potential to provide substantial improvements to interferometric observations of protoplanetary disks\cite{Yamaguchi2020SuperresolutionIO}. Moreover, estimates of relative dust optical depth and source brightness are also important\cite{harvey2012first}. Ultra-high-resolution IR images are also needed in the study of ongoing star formation\cite{Biller2005HighResolutionMI,Megeath1996EvidenceFO}. \rr{IRSR algorithms are also increasingly being deployed on embedded systems and edge computing platforms within vehicles, such as NVIDIA's Jetson series or specialized FPGAs\footnote{https://developer.nvidia.com/embedded/jetson-modules}. This necessitates the development of lightweight and efficient network architectures (e.g., using knowledge distillation or network pruning) to meet strict latency requirements for real-time decision-making. Companies like FLIR Systems (now Teledyne FLIR)\footnote{https://www.flir.eu/} provide thermal camera modules specifically for the automotive industry. The SR techniques implemented in these systems often prioritize speed and robustness over achieving the highest possible reference metrics, sometimes using simpler interpolation methods combined with fast neural enhancement filters rather than complex generative models.} Applications in other domains can be found in more detail in Tab. \ref{tab.1}.
% \section{The Design, Intent, and \\ Limitations of the Templates}
% The templates are intended to {\bf{approximate the final look and page length of the articles/papers}}. {\bf{They are NOT intended to be the final produced work that is displayed in print or on IEEEXplore\textsuperscript{\textregistered}}}. They will help to give the authors an approximation of the number of pages that will be in the final version. The structure of the \LaTeX\ files, as designed, enable easy conversion to XML for the composition systems used by the IEEE. The XML files are used to produce the final print/IEEEXplore pdf and then converted to HTML for IEEEXplore.

\section{Related Methods}

In this section, we will introduce IR image super-resolution methods, including traditional methods and deep learning-based methods. For the traditional algorithms, there will be three parts: frequency domain-based, dictionary-based, and other methods. Then, the deep learning methods will be presented.

\begin{figure*}[!t]
\centering
\includegraphics[width=\textwidth]{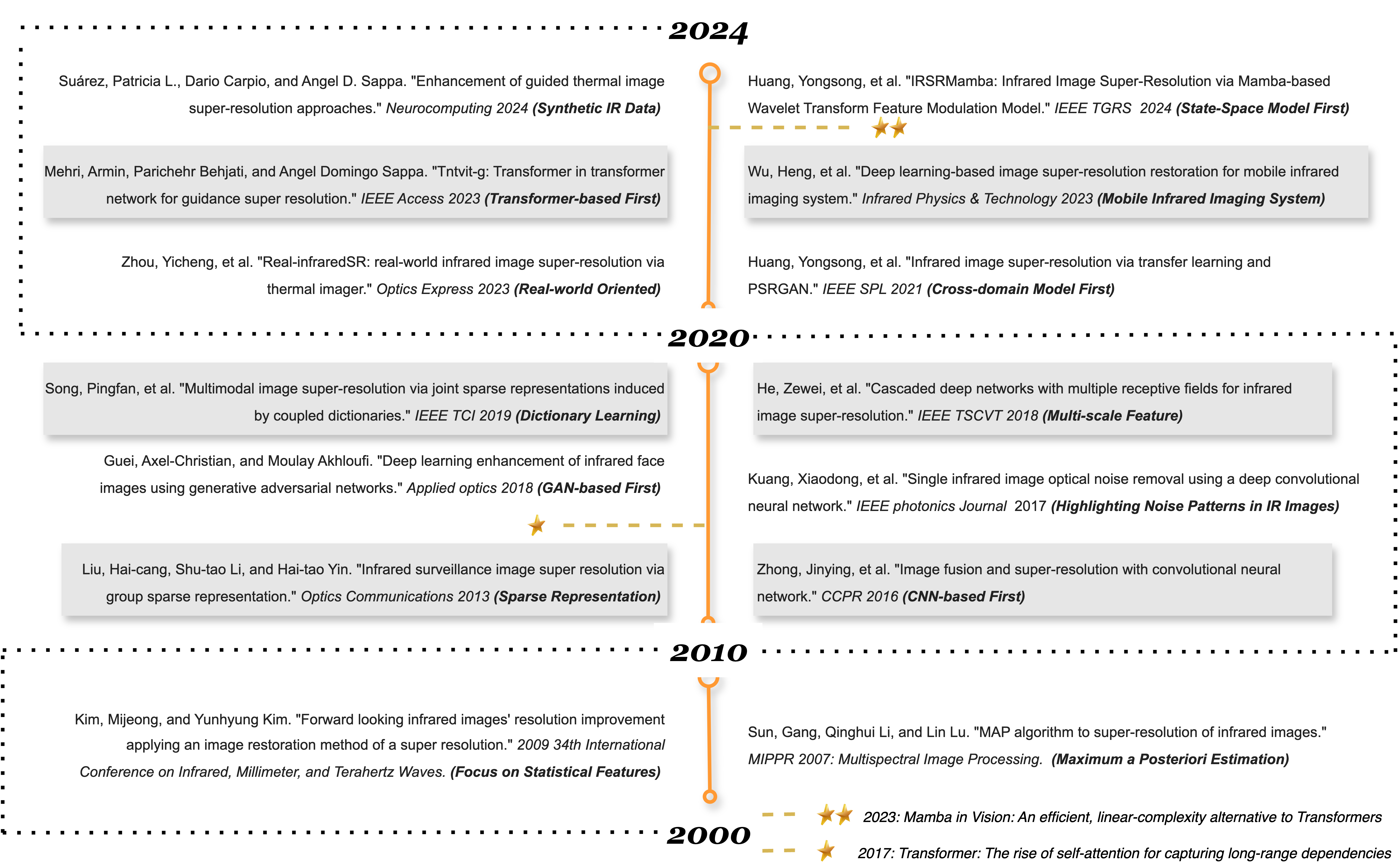}%
\label{fig_first_case}
\caption{\tcmi{The timeline of representative IRSR algorithms (from year 2000 to 2024).}}
\label{fig_timeline}
\end{figure*}

\tcmi{Fig.\ref{fig_timeline} provides a chronological overview of representative IRSR algorithms, illustrating the key advancements in this field from 2000 to 2024. The timeline highlights how traditional methods, such as frequency domain-based approaches and dictionary learning algorithms\cite{song2019multimodal}, laid the foundation for early IRSR research. These methods, prevalent from 2000 to 2010, primarily focused on mathematical models and handcrafted features to enhance IR image quality\cite{kim2009forward,sun2007map,liu2013infrared}. However, they often faced limitations in handling complex noise patterns\cite{kuang2017single} and adapting to diverse real-world scenarios. This paved the way for the adoption of deep learning approaches, starting in 2011, particularly CNN-based methods\cite{zhong2016image}. By learning hierarchical feature representations, these methods achieved significant improvements in reconstruction quality and robustness, marking a turning point in the development of IRSR techniques.}

\tcmi{In recent years, GANs and Transformer-based models have pushed the boundaries of IRSR methodologies even further. GANs, introduced around 2018, enabled realistic texture generation, effectively addressing the unique noise characteristics of IR images\cite{guei2018deep}. Meanwhile, Transformer-based models, exemplified by recent works such as \cite{mehri2023tntvit}, have enhanced global contextual modeling capabilities, resulting in superior performance in preserving structural consistency and fine details. This evolution reflects a broader trend toward leveraging advanced deep learning architectures to overcome the inherent challenges of IRSR. Additionally, recent studies have extended the scope of IRSR by exploring cross-domain learning\cite{huang2021infrared}, mobile infrared imaging systems\cite{wu2023deep}, and state-space modeling\cite{huang2024irsrmamba}, demonstrating the growing practical relevance and applicability of these techniques.}

\tcmi{Based on the advancements in IR imaging systems and deep learning techniques, researchers are increasingly focusing on developing more efficient image processing methods to enhance IR image quality further. As such, this survey categorizes image processing approaches into two primary types: traditional methods and deep learning-based techniques. To provide a clear comparison, Table \ref{tabcomparison} lists several representative works, summarizing their performance and contributions. This comparison not only highlights the progress made in the field but also underscores the evolving role of deep learning in addressing the specific challenges of IR image super-resolution.}

\begin{table*}[!t]
\centering
\caption{Performance comparison {[}in terms of PSNR (dB) and SSIM{]} over selected test dataset for 4 or 3 upscaling factors. Some of these works use visible images\cite{agustsson2017ntire} as training datasets, such as 7 and 9. The test datasets for these representative methods depend on the task, check the references for more details. \rr{$\oslash$ denotes that the result was not reported in the original paper or is not applicable under the current evaluation settings.}}
\label{tabcomparison}
\renewcommand\arraystretch{1.5}

\resizebox{0.8\textwidth}{!}{%
\begin{tabular}{@{}ccccc@{}}

\toprule
No. & \multicolumn{1}{c|}{Methods}                                                  & \multicolumn{1}{c|}{Year} & \multicolumn{1}{c|}{Ref. $\times$ 4 (PSNR/SSIM)} & Keywords                                    \\ \midrule
1   & Wu, Wenhao, et al. \cite{wu2022meta}                          & 2022                      & 31.69/0.7877                                & Meta-learning; Lightweight network          \\
\rowcolor[HTML]{EFEFEF} 
2   & Huang, Yongsong, et al.\cite{huang2021infrared}               & 2021                      & 33.13/0.8282                                & Transfer learning;  Small sample            \\
3   & Liu, Qing-Ming, et al. \cite{liu2021infrared}                 & 2021                      & 32.23/0.8720                              & Attention mechanism; GAN                    \\
\rowcolor[HTML]{EFEFEF} 
4   & Prajapati, Kalpesh, et al.\cite{prajapati2021channel}         & 2021                      & 34.90/0.9134                              & Attention mechanism;  Channel splitting     \\
5   & Marivani, Iman, et al. \cite{marivani2020joint}               & 2020                      & 35.19/0.9888                               & Sparse coding; Image fusion                 \\
\rowcolor[HTML]{EFEFEF} 
6   & Marivani, Iman, et al. \cite{marivani2020multimodal}          & 2020                      & 34.49/0.9853                              & Multimodal image; Residual learning         \\
7   & Yao, Tingting, et al.\cite{yao2020infrared}                   & 2020                      & 34.54/0.8807                               & Dictionary learning; CNN                    \\
\rowcolor[HTML]{EFEFEF} 
8   & Song, Pingfan, et al.\cite{song2019multimodal}                & 2020                      & 36.36/0.9796                               & Dictionary learning; Sparse representations \\
9  & Batchuluun, Ganbayar, et al. \cite{batchuluun2020thermal}     & 2020                      & 22.99/0.9760                              & Image deblurring; GAN                       \\
\rowcolor[HTML]{EFEFEF} 
10  & Rivadeneira, Rafael E, et al.\cite{rivadeneira2019thermal}    & 2019                      & 37.85/$\oslash $                                     & Thermal images; CNN                         \\
11  & Suryanarayana, Gunnam, et al.\cite{suryanarayana2019infrared} & 2019                      & 31.40/0.9513                                & Discrete wavelet transform; CNN             \\
\rowcolor[HTML]{EFEFEF} 
12  & Marivani, Iman, et al. \cite{marivani2019multimodal}          & 2019                      & 33.19/$\oslash $                                     & Sparse coding; CNN                          \\
13  & He, Zewei, et al.\cite{he2018cascaded}                        & 2018                      & 36.02/0.9230                                & Multiple receptive fields; CNN              \\
\rowcolor[HTML]{EFEFEF} 
14  & Sun, Chao, et al.\cite{sun2018rapid}                          & 2018                      & 32.71/$\oslash $                                     & Zoom mechanism; Transfer learning           \\
15  & Han, Tae Young, et al.\cite{han2018infrared}                  & 2018                      & 39.69/0.9582 ($\times$ 3)                           & Frequency components; CNN                   \\
\rowcolor[HTML]{EFEFEF} 
16  & Zhao, Yao, et al. \cite{zhao2016learning}                     & 2016                      & 30.69/0.9031 ($\times$ 3)                           & Compressed sensing; Dictionary learning;    \\ \bottomrule
\end{tabular}%
}
\end{table*}
\vspace{-0.5cm}

\subsection{Traditional Methods\label{sec3.1}}

In traditional image processing algorithms, the IR image super-resolution follows the direction of visible image super-resolution. Specifically, these approaches include \textbf{frequency domain-based, dictionary-based, and other methods.}

\subsubsection{Frequency Domain-based}

\rr{Frequency-domain analysis is well-suited for IRSR due to the unique spectral properties of thermal imagery. Unlike their visible-light counterparts, IR images are inherently low-pass, characterized by weak edge contrast and sparse high-frequency details. This is a direct consequence of the underlying physics: thermal diffusion naturally smooths temperature gradients, while the long-wavelength nature of IR radiation leads to diffraction effects that attenuate fine textures. This dominance of low-frequency information makes methods that explicitly separate and model spectral components a powerful and principled approach for reconstruction.}

For the frequency domain-based approach, we will decompose the IR image into two domains: the air domain, and the frequency domain\cite{wang2009analysis}. Especially for multiple complex frequency domain components, it is often not included by a simple FFT. For example, there is a significant difference between high-frequency and low-frequency information in IR images. \textit{Choi et al.} propose a new frequency-domain processing method for this problem: firstly, the edge pixels are distinguished, then the high-frequency information is enhanced independently, and finally, a high-resolution image is available\cite{choi2011resolution}. 

\begin{eqnarray}
e(\hat{k}, l) & = & \left(1-e^{-\frac{\left(\lambda_{1}+\lambda_{2}\right)}{\sigma_{\mathrm{sum}}}}\right)\left(1-e^{-\frac{\left(\lambda_{1}+\varepsilon\right) /\left(\lambda_{2}+\varepsilon\right)}{\sigma_{\text {ratio }}}}\right) 
\label{eq.4}
\end{eqnarray}

As shown in Eq. \ref{eq.4}, the researchers introduced prior knowledge skillfully. Two eigenvalues are denoted by $\lambda_1$ and $\lambda_2$, where $\lambda_1 \geq \lambda_2 \geq 0$, which are recalculated from the structure tensor at pixel $(\hat{k}, l)$, $\sigma_{\text {sum }}$ and $\sigma_{\text {ratio }}$ indicate established parameters. In order to avoid division by zero, a tiny constant named $\varepsilon$ is created. It is important to mention that we will need a carefully engineered prior with expert knowledge to ensure the performance, which is difficult to achieve with the current relatively weak physical modeling \cite{park2003super}. On the other hand, there are also some frequency domain-based methods focusing on the reconstruction time issue and aiming to accelerate\cite{mao2016infrared}. Because we will fuse 4 images to get better image resolution. In addition, since the edges of the original image are bold, it will be difficult to match the edges during registration.

\subsubsection{Dictionary-based}

Considering the good interpretability, the dictionary-based reconstruction methods have gained a lot of popularity among researchers in IR images. The formalization of the dictionary-based reconstruction method can be found in this work\cite{chang2004super}. In summary, the dictionary-based approach focuses on patches that build a bridge between $I_{HR}$ and $I_{LR}$. These patches include features such as \textbf{1)} correspondence, \textbf{2)} the number of patches, and \textbf{3)} inter-patch relationships. The correspondence relationship and the number of patches are used to ensure that the mapping between the $I_{LR}$ and the $I_{HR}$ is complete. And the patterns of the low-resolution image can be preserved in the high-resolution image by the inter-patch relationship. The proposed methods improve the image quality from these perspectives, and more details are shown in Tab. \ref{tab.2}.

We will briefly describe the representative dictionary-based works: First, dividing the groups. On the one hand, it has been observed that the similarity between image groups impacts the reconstruction, that is, similarity and compatibility as previously discussed. This approach considers group similarity and proposes positional constraints. The position between groups is used as a feature to constrain the reconstruction of the image\cite{cheng2014infrared}. In addition, multiple information (multi-view) is introduced to reconstruct images using more groups to represent instead of isolated groups\cite{yang2015infrared,yang2017multi}. Researchers use multi-scale groups to compile dictionaries, which can provide rich scale information and diverse structural information\cite{yang2017multi}. In specific tasks, such as pedestrian recognition, if the person as a target is segmented into different blocks will hurt the recognition results. In this situation, multi-scale blocks can keep the person retained in the complete block while further constructing the dictionary\cite{zou2018super}. Second, the relationship between the total number of dictionaries and reconstruction. The researchers dropped the question that the single dictionary created from IR images has limited the development of dictionary-based methods or not. This limitation is expressed in two directions: fewer dictionary pairs and less information about the IR images. It is proposed that a single dictionary cannot represent all the structural information. And, multiple dictionaries should be used to support the reconstruction\cite{yang2016fast}. 

\begin{equation}
\begin{aligned}
\mathbf{x}^h &=\boldsymbol{\Psi}_c^h \mathbf{z}+\boldsymbol{\Psi}^h \mathbf{u}, \\
\mathbf{x}^l &=\boldsymbol{\Psi}_c^l \mathbf{z}+\boldsymbol{\Psi}^l \mathbf{u}, \\
\mathbf{y} &=\boldsymbol{\Phi}_c \mathbf{z}+\boldsymbol{\Phi} \mathbf{v}.
\end{aligned}
\label{eq.5}
\end{equation}

Moreover, it is also the first time that extra dictionaries are constructed using visible image information obtained from the same scene in a dictionary-based approach to help the reconstruction. Third, the information, from other types of images, can also be used to support the dictionary construction. Because the image information in the IR image itself is insufficient due to the poor imaging environment. Then, people started to focus on introducing extra information. This algorithm is developed to enhance the resolution of the target LR image with the aid of another guidance HR image modality\cite{song2019multimodal}. More details are shown in Eq. \ref{eq.5}.

They express the LR image patch $\mathbf{x}^l \in \mathbb{R}^M$ and $\mathrm{HR}$ image patch $\mathbf{x}^h \in \mathbb{R}^N$ of the same image modality, and the guidance $\mathrm{HR}$ patch of another different image modality $\mathbf{y} \in \mathbb{R}^N$. The definitions of the symbols can be found in Abbreviations and Notations. Some works noticed that gradient information of IR images would be beneficial for reconstruction\cite{wang2021research,chen2018infrared}.
% Further, the joint sparse representation induced by the coupling dictionary is used to capture the complex dependencies between different modalities in the sparse feature domain.

\begin{table*}[t]
\centering
\caption{Dictionary-based reconstruction methods.}
\label{tab.2}
\renewcommand\arraystretch{1.2}
\resizebox{0.9\textwidth}{!}{
\begin{tabular}{@{}ccc@{}}
\toprule
\textbf{Category} &
  \textbf{Method} &
  \textbf{Key idea} \\ \midrule
\multicolumn{1}{c|}{Patches} &
  \multicolumn{1}{c|}{Cheng-Zhi, Deng, et al. (2014)\cite{cheng2014infrared}} &
  \begin{tabular}[c]{@{}c@{}}Considering the similarity between groups, the locality-constrained group is \\ proposed to guide the reconstruction by using the positions between groups as features.\end{tabular} \\ \cmidrule(l){2-3} 
\multicolumn{1}{c|}{} &
  \multicolumn{1}{c|}{\begin{tabular}[c]{@{}c@{}}Yang, Xiaomin, et al. (2015)\cite{yang2015infrared}\\ Yang, Xiaomin, et al. (2017)\cite{yang2017multi}\end{tabular}} &
  Introduce multi-view information and multiscale information to build patches. \\ \midrule
\multicolumn{1}{c|}{Dictionary} &
  \multicolumn{1}{c|}{Yang, Xiaomin, et al. (2016)\cite{yang2016fast}} &
  Using more dictionaries. \\ \cmidrule(l){2-3} 
\multicolumn{1}{c|}{} &
  \multicolumn{1}{l|}{Yao, Tingting, et al. (2020)\cite{yao2020infrared}} &
  \begin{tabular}[c]{@{}c@{}}It is proposed to capture more IR image features with neural networks and then use \\ this information to build dictionaries.\end{tabular} \\ \midrule
\multicolumn{1}{c|}{\begin{tabular}[c]{@{}c@{}}Correspondence \\ Relationship\end{tabular}} &
  \multicolumn{1}{c|}{Zou, Erbo, et al. (2018)\cite{zou2018super}} &
  \begin{tabular}[c]{@{}c@{}}To better represent the pattern of groups, \\ this work improves performance by learning the correspondence several times.\end{tabular} \\ \midrule
\multicolumn{1}{c|}{Extra Information} &
  \multicolumn{1}{c|}{Song, Pingfan, et al. (2019)\cite{song2019multimodal}} &
  \begin{tabular}[c]{@{}c@{}}Other image modalities, such as RGB images, are introduced to enhance edge \\ and structure primitives information.\end{tabular} \\ \cmidrule(l){2-3} 
\multicolumn{1}{c|}{} &
  \multicolumn{1}{c|}{\begin{tabular}[c|]{@{}c@{}}Chen, Zuming, et al. (2018)\cite{chen2018infrared}\\ Wang, Yan, et al. (2021) \cite{wang2021research}\end{tabular}} &
  These works propose to pay more attention to the gradient information in IR images. \\ \bottomrule
\end{tabular}
}
\end{table*}

\subsubsection{Other Traditional Algorithms}
There are other traditional algorithms used for IR image super-resolution listed here, as shown in Tab.\ref{tab.3}. Some methods aim to transfer the super-resolution methods in visible images to IR images. \textbf{1) Iterative} methods are used in some biometric application situations, such as iris recognition\cite{alonso2017iris}, and laser line scanning thermography\cite{ahmadi2020super}. However, such methods are not fine-tuned according to the special patterns in IR image. \textbf{2) Sparse representations} are used in land remote sensing where the IR term with IR information is calculated by computing the normalized difference target index (NDTI). There are also swarm optimization algorithms applied to minimize the terms to get the land cover target mapping (LCTM)\cite{wang2022compressed}. \textbf{3) In sparsity-based methods,} researchers pay more attention to the specific patterns in IR images. In such methods, it is still common to process information in the frequency domain. However, if we only consider the frequency domain, the contribution from the space domain information may be ignored. For example, the similarity between neighboring points is considered, different from the general frequency domain-based approach, where the image gradient represented by sparsity is mentioned\cite{liu2019infrared,gungor2019transform}. In addition, there are new directions, \ie, using higher-order image derivatives to receive the image prior to sparsity\cite{liu2018infrared,chen2019research}.

\textit{S. C. Park et al.} described the details about the projection method in his work carefully\cite{park2003super}.
Such methods require appropriate priors to guide the projection, and researchers considered the phase characteristics in images to improve the algorithm\cite{zhang2019super}. It is also proposed to introduce illumination sensitivity prior to construct iterations\cite{liu2016improved}.

On the other hand, regularization methods also play a key role. \textit{Yu Hui et al.}\cite{hui2014novel} believed that the regularization term can also suppress the unnecessary low-frequency components and enhance the image quality. A two-step fully-variable sparse reconstruction algorithm is designed by \textit{Wang et al.}\cite{wang2022compressed}  The fully-variable regularization term is added to the classical compressed perceptual model. 

The weaknesses in such methods are also easily observed in experiments: \textit{Dijk et al.} mention that regularization methods influence the performance in detection tasks\cite{dijk2009performance}. Moreover, it is supposed that the model is not considered for complex nonlinear degradation factors. The result is that algorithms have limited fitness for real-world applications\cite{yu2013single}.

\begin{table}[!t]
\centering
\caption{IR image super-resolution via other traditional algorithms.}
\label{tab.3}
\renewcommand\arraystretch{1.5}
\begin{tabular}{@{}cc@{}}
\toprule
\multicolumn{1}{c|}{\textbf{Category}} & \textbf{Method }                                                                                             \\ \midrule
\rowcolor[HTML]{EFEFEF} 
Iterative                     & \begin{tabular}[c]{@{}c@{}}A. Fernandez, et al. (2017) \cite{alonso2017iris}\\ Ahmadi, Samim, et al. (2020)\cite{ahmadi2020super}\end{tabular} \\
Sparse Representations        & Wang, Peng, et al. (2021) \cite{wang2021land}                                                                          \\
\rowcolor[HTML]{EFEFEF} 
Sparsity &
  \begin{tabular}[c]{@{}c@{}}Liu, Xingguo, et al. (2019)\cite{liu2019infrared}\\ Güngör, Alper, et al. (2019) \cite{gungor2019transform}\\ Liu, Xingguo, et al. (2018)\cite{liu2018infrared}\\ Chen, Shaojun, et al. (2019)\cite{chen2019research}\end{tabular} \\
Projection                    & \begin{tabular}[c]{@{}c@{}}Zhang, X. F., et al. (2019)\cite{zhang2019super}\\ Liu, Jinsong, et al. (2016)\cite{liu2016improved}\end{tabular}   \\
\rowcolor[HTML]{EFEFEF} 
Regularization &
  \begin{tabular}[c]{@{}c@{}}Yu, Hui, et al. (2013)\cite{yu2013single}\\ Hui, Yu, et al. (2014)\cite{hui2014novel}\\ Wang, Yan, et al. (2022)\cite{wang2022compressed}\end{tabular} \\ \bottomrule
\end{tabular}
\end{table}

\begin{figure}[!t]
\centerline{\includegraphics[width=\columnwidth]{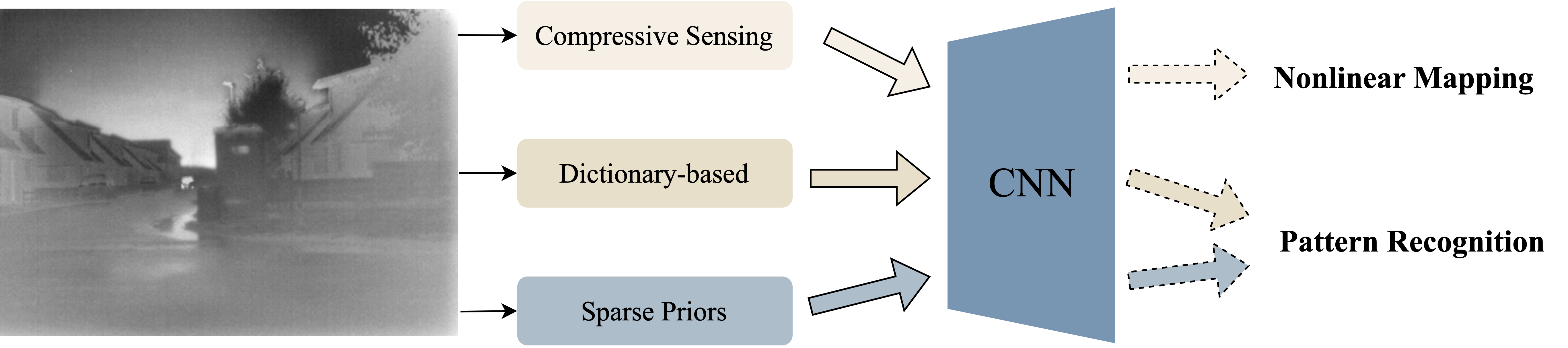}}
\caption{CNNs contribute to other traditional methods used to learn nonlinear mappings and pattern recognition.}
\label{fig4}
\end{figure}

\begin{table}[h]
\centering
\renewcommand\arraystretch{1.5}
\caption{CNN transfers on traditional methods.}
\label{tab.4}
\resizebox{\columnwidth}{!}{
\begin{tabular}{@{}ccc@{}}
\toprule
\multicolumn{1}{c|}{\textbf{Category}}                         & \multicolumn{1}{c|}{\textbf{Method}} & \textbf{Highlight} \\ \midrule
\rowcolor[HTML]{EFEFEF} 
\begin{tabular}[c]{@{}c@{}}Compressive \\ Sensing\end{tabular} & Zhang, Xudong, et al. (2018)\cite{zhang2018infrared}         & Nonlinear Mapping  \\
Dictionary-based & Yao, Tingting, et al. (2020)\cite{yao2020infrared}  &                                       \\
Sparse Priors    & Marivani, Iman, et al. (2020)\cite{marivani2020joint} & {Pattern Recognition} \\ \bottomrule
\end{tabular}
}
\end{table}

\subsection{Deep Learning Methods}

With the development of deep learning, neural networks have shown remarkable performance in image super-resolution with powerful fitting ability\cite{dong2015image,dong2016accelerating,kim2016accurate,lim2017enhanced,wang2018esrgan,ledig2017photo}. However, few previously published studies have been able to draw on any systematic research into IR image-specific patterns. These unique patterns include gradient information, high or low-frequency information, extra information, etc\cite{huang2021infrared,huang2021pricai,fan2021infrared,zhang2019super,liu2022infrared}.  In this section, we will also focus on the differences between IR images and visible images in the deep learning field of super-resolution. In summary, the challenges of IR image super-resolution in deep learning include specific patterns, difficulty in representing patterns, and poor image quality. More details will be shown in the next section.

\subsubsection{CNN-based}
There are two major trends for CNN-based models in IR image super-resolution: \textbf{1)} First, researchers introduced CNNs as contributors to improving traditional algorithms (see Fig. \ref{fig4}, Tab.\ref{tab.4}). \textbf{2)} Further, it became popular to use CNNs to build end-to-end models used in IR image super-resolution.

\textbf{CNN \& Traditional algorithms:} As mentioned in Sec .\ref{sec3.1}, traditional methods require enough prior knowledge to achieve image super-resolution. However, weak mathematical analysis limits the proposed algorithms. CNN can learn nonlinear mappings without indication because of its powerful fitting ability\cite{dong2015image,dong2016accelerating,kim2016accurate,lim2017enhanced,wang2018esrgan,ledig2017photo}. It allows people to free themselves from the task to seek for priors. Initially, neural networks were not used to reconstruct IR images directly. Rather, it was carefully used to help traditional algorithms improve performance through nonlinear mapping and pattern recognition.

\textit{1) Nonlinear mapping:} In infrared camera systems, there will be little high-frequency information and low-frequency information in the captured infrared image compared to the visible image due to the poor imaging environment (see Sec.\ref{sec.2.1}). It becomes a challenge to better represent these two kinds of information in IR images super-resolution. Then, the researchers proposed that CNN can be used to represent the nonlinear mapping of low and high frequency information in the latent space. In \textit{Zhang's work}\cite{zhang2018infrared}, the SR image is reconstructed using compressed sensing first. Then the SR image and HR image are reduced to receive high-frequency noise information. Finally, the information is fed into CNN to learn nonlinear mapping.

\begin{equation}
\mathcal{L}=\frac{1}{2 N} \sum_{i=1}^N\|\widetilde{R}(\Theta, i)-R(\Theta, i)\|^2
\label{eq.6}
\end{equation}

Eq.\ref{eq.6} denotes the loss function to learn the trainable parameters $\Theta$ in $\mathrm{CNN}$. Corresponding to $i$ th training image, $\widetilde{R}(\Theta, i)$ represents the estimated residual image produced by CNN, while $R(\Theta, i)$ represents the true residual image used for training\cite{zhang2018infrared}. Experiments show that this approach using CNNs to represent nonlinear mappings for reconstructing IR images can enhance detailed information.

\textit{2) Pattern recognition:} On the other hand, \textit{Yao,  Tingtin, et al.} \cite{yao2020infrared} proposed to use CNN as a feature extractor in the dictionary-based approach. This method also focuses on the problem that the detailed information in IR images is important, and CNN can build the dictionary with more details added to the representation. Novel interpretable operators are proposed to construct the basic module as the sparse representation for prior extraction, another module to represent the edge information of visible images, and finally fused in a residual network using skip connection\cite{marivani2020joint}.

\begin{figure}[!t]
\centerline{\includegraphics[width=\columnwidth]{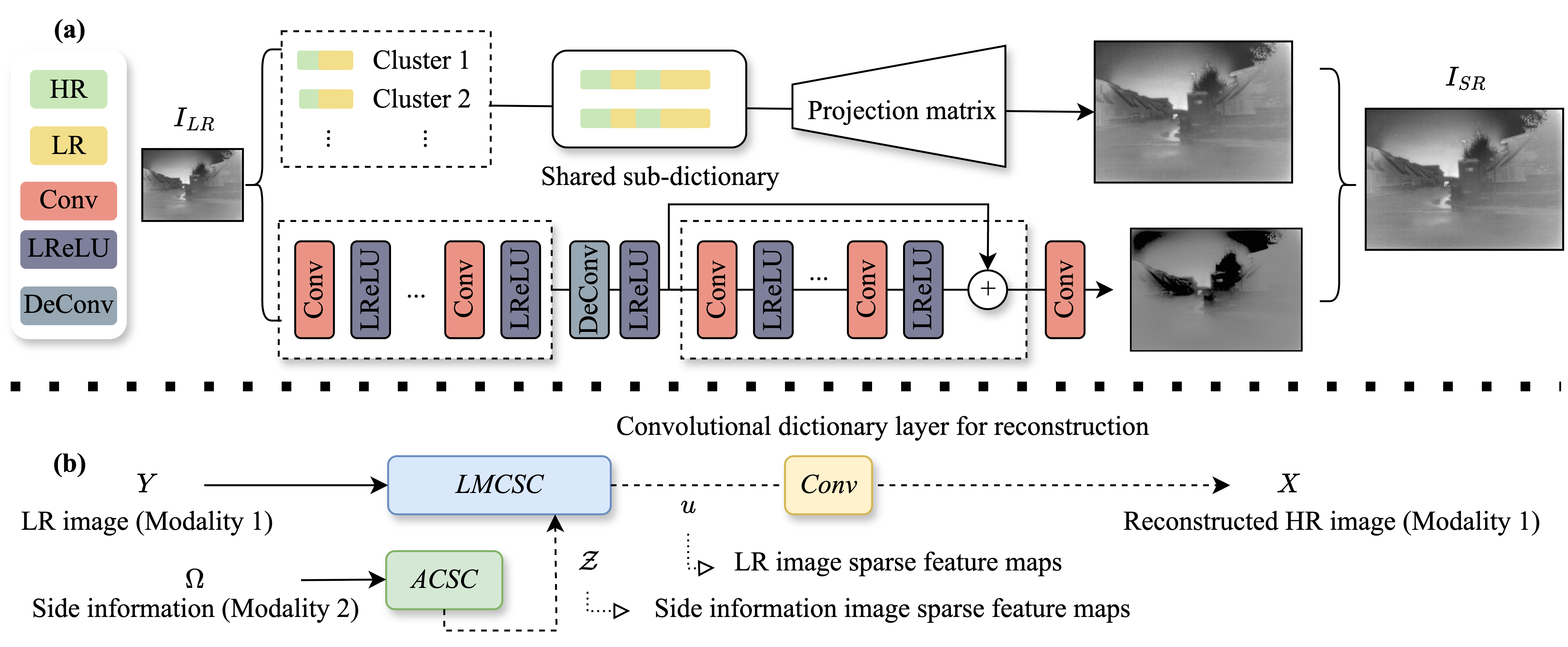}}
\caption{(a) shows the CNN used as a feature extractor in a dictionary-based approach; (b) is an illustration for the CNN used to represent prior.}
\label{fig5}
\end{figure}

More details are shown in Fig.\ref{fig5}. For (b), $X$ and $Y$ denote the LR image, and reconstructed SR image, respectively. The convolutional sparse codes $\mathcal{Z}$ of the guidance $\mathrm{HR}$ image $\boldsymbol{\Omega}$ are similar to $u$ by means of the $\ell_1$-norm according to the literature\cite{marivani2020joint}. Furthermore, residual networks with multiple branches are used to reconstruct high and low-frequency information, respectively\cite{yang2020deep}.

Beyond introducing CNNs, these existing methods started to introduce visible images to guide the reconstruction of IR images purposefully, which represents an entirely new direction. Further, attention has been focused on image characteristics, specifically the difference between high and low-frequency information in IR images. Details will be described in the next sections.

\textbf{CNN \& End-to-end models:} In this section, the end-to-end models used in IR image super-resolution will be presented. The high-frequency and low-frequency information in IR images is responsible for the image outline and edges, respectively. Moreover, the IR image has less edge information compared to the visible image because of the imaging system (see Sec.\ref{sec.2.1}). First, the researchers propose that the independent extraction for high-frequency and low-frequency information in IR images is achieved by CNN. In \cite{zou2021super}, \textit{Zou et al.} used residual networks to build a model similar to U-Net. The multi-receptive field module in this work is supposed to help represent the features of high and low-frequency information. Similar work is \cite{fan2021infrared}, and the residual network is also used. However, the information distillation approach is used in the backbone network, which was first proposed in \cite{huang2021infrared} to benefit IR image super-resolution. This Information distillation-based model is also considered to be helpful for different information extraction. In addition, \textit{Kwasniewska et al.} used a wide receptive field residual network constructed by dense connection\cite{kwasniewska2020super}. According to the experimental results, this work confirmed that wide receptive fields are effectively used for low-contrast images. \textit{Prajapati, Kalpesh, et al.} suggest that the common residual design allows for retaining too many redundant features in the network. A new module for integration with multiple attention mechanisms is proposed in their work. According to the experimental results, this novel attention mechanism module will help to enhance the high-frequency information representation\cite{prajapati2021channel}. Finally, not only the residual network is used, but also the importance of edges is observed\cite{gao2022maritime}. The method that enhances high-frequency details and contacts information-tail modules to improve details is the highlight. 

% More details are shown in Fig.\ref{fig6}. And $F_{n l m^D}$ is the output of the previous (\ie $D$ ) recursive block

\begin{figure}[!t]
\centerline{\includegraphics[width=0.7\columnwidth]{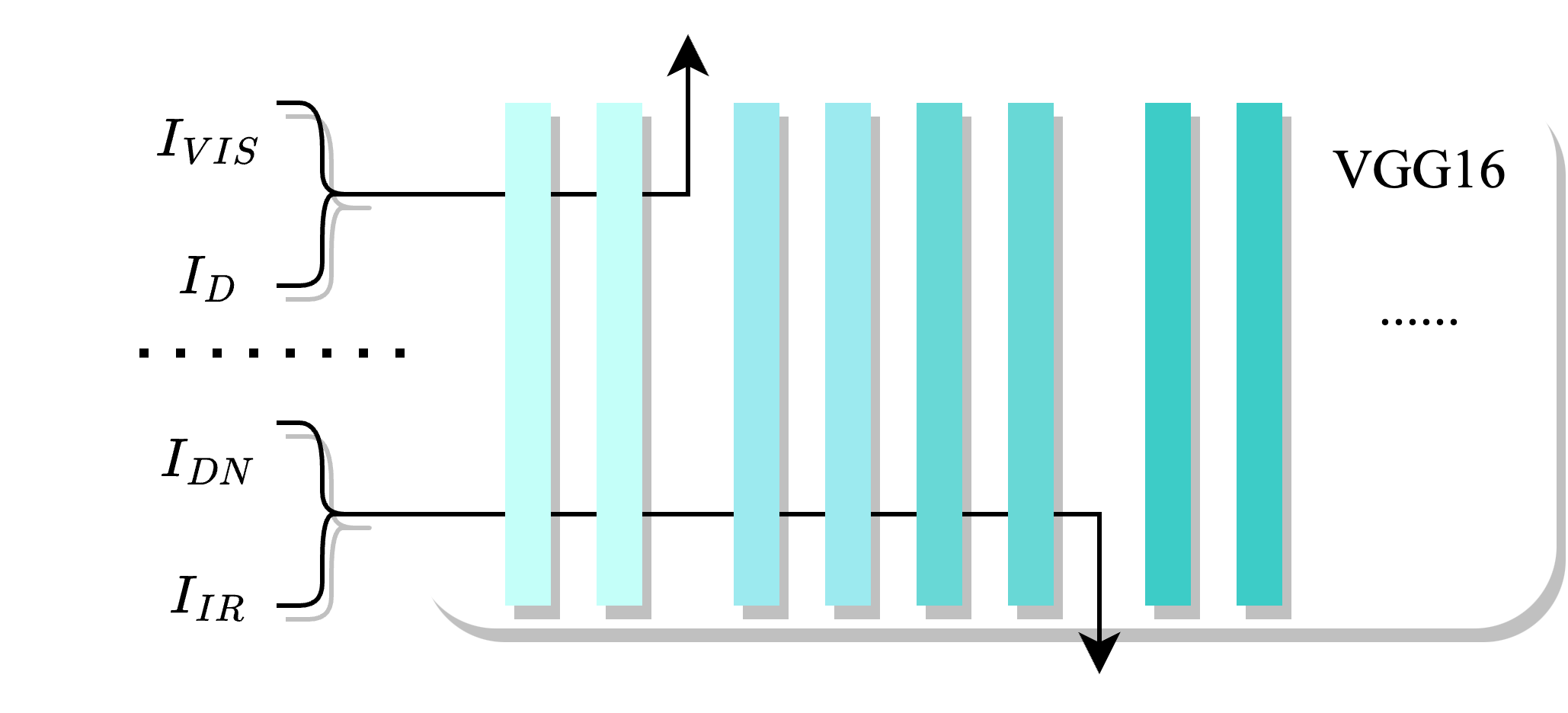}}
\caption{$I_{DN}$ denotes the denoised output, while $I_{D}$  is the output from the author's proposed network\cite{yang2020deep}.}
\label{fig7}
\end{figure}

% \begin{figure*}[!t]
% \centering
% \includegraphics[width=\textwidth]{fig/fig6.png}%
% \label{fig_first_case}
% \caption{In (a), the researchers used a network similar to U-Net to extract high and low-frequency information independently\cite{zou2021super}. For the same purpose, (b) demonstrates that the information distillation method can also be used\cite{kwasniewska2020super}. Finally, the module in (c) shows that wide receptive fields are beneficial for low-contrast images\cite{gao2022maritime}.}
% \label{fig6}
% \end{figure*}

\begin{table}[!t]
\centering
\caption{Ablation studies for introducing visible image information. From 1 to 4 stars, indicates that the subjective evaluation is improving.}
\label{tab5}
\begin{tabular}{@{}c|cccc@{}}
\toprule
Residual Network                  & $\surd$                      & $\surd$                              & $\surd$                                      &                                 \\ \cmidrule(r){1-1}
Visible Images (Training Dataset) &                              & $\surd$                              & $\surd$                                      & $\surd$                         \\ \cmidrule(r){1-1}
Novelty Loss                      &                              &                                      & $\surd$                                      & $\surd$                         \\ \cmidrule(r){1-1}
Novelty Network                   &                              &                                      &                                              & $\surd$                         \\ \midrule
Results                           & \multicolumn{1}{c|}{$\star$} & \multicolumn{1}{c|}{$\star$ $\star$} & \multicolumn{1}{c|}{$\star$ $\star$ $\star$} & $\star$ $\star$ $\star$ $\star$ \\ \bottomrule
\end{tabular}
\end{table}

The approaches mentioned above show that the two variables in the latent space (low-frequency and high-frequency information) are treated independently. Then novel methods are used to represent the nonlinear mapping between these two independent variables respectively. In the next section, we will describe another category of methods that introduce extra information: visible image information (see Tab.\ref{tab5}).

For introducing visible image information, the researchers focused on the following components: dataset, network structure, and loss function. It is natural to associate using visible images to train neural networks because they are cheap and easy to use, with rich detailed information. However, it is difficult to fit the data distribution for IR images with the model trained on the visible image dataset due to the domain transfer challenge\cite{huang2021infrared,rivadeneira2019thermal}. In other words, the SR images have poor quality. Then, adaptations to the loss function are proposed. As we all know, the loss function is important for optimizing neural networks. \textit{Patel, Heena M., et al.} proposed that a combined loss function, with two terms for L1 and SSIM loss, could be used\cite{patel2021thermisrnet}. According to the ablation experiments, this combined loss function would be beneficial for their proposed improved dense blocks. On the other hand, it also inspired that appropriate loss functions would be useful for introducing information from visible images In \cite{yang2020deep}, the authors propose that the perceptual similarity between IR images $I_{IR}$ and paired visible images $I_{VIS}$ can be captured by CNNs. And there are features present in the middle layer of the neural network and they can be used as loss functions (see Fig.\ref{fig7}).

Next, we will introduce novel network structures for using visible image patterns. \textit{Zou, Yan, et al.} proposed a dual-path residual network that directly fuses the features from visible and IR images in the channel\cite{zou2020infrared}. Further, \textit{Oz, Navot, et al.} consider that dimensionality compression will influence the performance in feature fusion. Because the correlation between adjacent pixels is ignored. Then, the dimensional change network for increasing channels was proposed\cite{oz2020rapid}.

\begin{figure}[t]
\centerline{\includegraphics[width=\columnwidth]{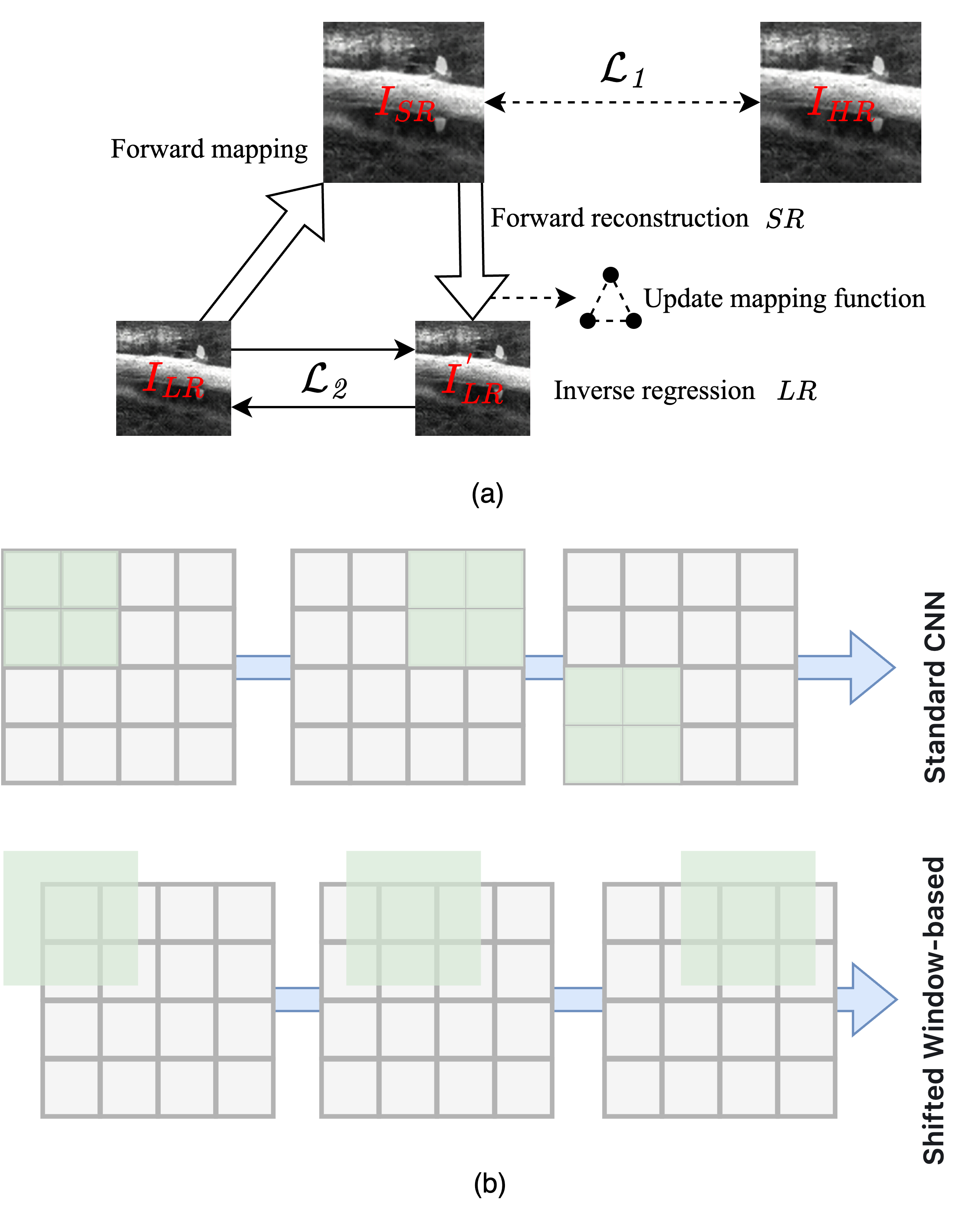}}
\caption{\rr{(a) The regression network structure used for IR image super-resolution. (b) Conceptual comparison of receptive field mechanisms. Top: A standard CNN employs a small, local kernel that slides across the feature map, limiting its ability to directly model long-range dependencies. Bottom: A shifted window-based Transformer computes attention within larger windows that are shifted between layers, enabling efficient global information exchange and the modeling of long-range dependencies.}}
\label{fig8}
\end{figure}

Finally, we will also present models related to super-resolution in multi-vision tasks. In many vision tasks\cite{zhong2016image,wang2022multimodal}, such as segmentation and fusion, super-resolution is applied to help other subjects as a pre-processing task. In particular, the regression network proposed in \cite{wang2022multimodal} is dedicated to prevent the irrelevant function mapping space on the reconstructed images by using forward generation and backward regression. This double mapping constraint is used for IR image super-resolution, and then the output that has better image quality is used in the fusion task. More details can be found in Fig.\ref{fig8}.

The loss function used in this work is: $x_i$ and $y_i$ respectively represent the LR image $I_{LR}$ and output SR images $I_{SR}$. $\mathcal{L}_{1}\left(F\left(x_i\right), y_i\right)$ and $\mathcal{L}_{2}\left(D\left(y_i\right), x_i\right)$ describe the loss functions of forward regression and inverse regression tasks, respectively. 

\begin{equation}
\mathcal{L} =\sum_{i=1}^N \mathcal{L}_{1}\left(F\left(x_i\right), y_i\right)+\lambda\mathcal{L}_{2}\left(D\left(y_i\right), x_i\right)
\label{eq.7}
\end{equation}

Moreover, this work mentions that intensity distribution and gradient information can represent thermal radiation and structural information, separately.

% \begin{figure}[t]
% \centerline{\includegraphics[width=\columnwidth]{fig/fig8.png}}
% \caption{Regression network structure.}
% \label{fig8}
% \end{figure}

In summary, the approaches in IR images and CNNs have first tried to combine the pattern recognition capabilities of CNNs into traditional reconstruction methods, such as dictionary reconstruction and sparse coding. Then started to reconstruct IR images directly using end-to-end models. Many module-based improvement methods were proposed, such as multi-scale information extraction. However, the high and low-frequency information in IR images have a significant gap. To use this information, the model structure has been improved further, for example, by presenting the information separately in different modules or using more complex modules. Compared to IR images, visible images have information with rich details. For this reason, researchers have started to try to use visible images to help reconstruct IR images. It includes designing new network structures and using strategies such as transfer learning. The purpose is to help neural networks can introduce more patterns and information from visible images.

\subsubsection{GAN-based}

After GAN\cite{goodfellow2020generative} was proposed, research based on adversarial training model is emerging in the super-resolution field. SRGAN\cite{ledig2017photo}, ESRGAN\cite{wang2018esrgan}, and various other types of GAN models\cite{ma2020structure} moved the field forward together through model improvement and mathematical analysis (WGAN\cite{gulrajani2017improved}) modifications. The same attention has been focused on the application and research of GAN models in the IR image super-resolution.

Initially, GAN models using normal images have been directly used in IR image reconstruction. Researchers\cite{shao2018single} used SRGAN straightly to reconstruct IR images and received reconstructed images that were quite acceptable. However, the blurred edges and unclear details are still a challenge. \cite{guei2018deep} used a modified DCGAN to reconstruct IR images, but the experimental results were not compared between the same category GAN methods. For this reason, it is difficult to describe the actual effectiveness.  On the other hand, the work from \cite{zhang2020single} compared the existing SR algorithms cross-sectionally on IR images. The experimental results show that the Super-Resolution Feedback Network (SRFBN)\cite{li2019feedback} model has the best universalization ability. But, for the GAN model, there are always unpleasant artifacts due to the model collapse. All these works illustrate the domain transfer difficulties possibly suffered by algorithms using visible images that are used in the IR images super-resolution task. Then, people started to consider the possibility of designing GAN models specifically for IR images. These studies are divided into the following categories: module improvement, introduce extra information.

\textit{1) Module improvement:} It is natural first to consider that the improvement method is to explore new modules to improve performance. \textit{Rivadeneira et al.\cite{rivadeneira2022novel}} uses the CycleGAN structure and employs ResNet as the generator's encoder. The self-attentive module is applied in the encoder and a new loss function is also proposed (see Eq. \ref{eq.8}). According to the discussion in this work, the Sobel edge detector can capture the contour consistency between the input image and the cyclic-generated image by calculating the mean squared difference between the images.

\begin{equation}
\scalebox{0.8}{% 缩小比例为 0.8
$\mathcal{L}_{\text {Sobel }}=\frac{1}{N} \sum_i\left\|\operatorname{Sobel}\left(G_{H 2 L}\left(G_{L 2 H}\left(I_L\right)\right)\right)-\operatorname{Sobel}\left(I_L\right)\right\|$
}
\label{eq.8}
\end{equation}

\begin{figure}[!t]
\centerline{\includegraphics[width=0.8\columnwidth]{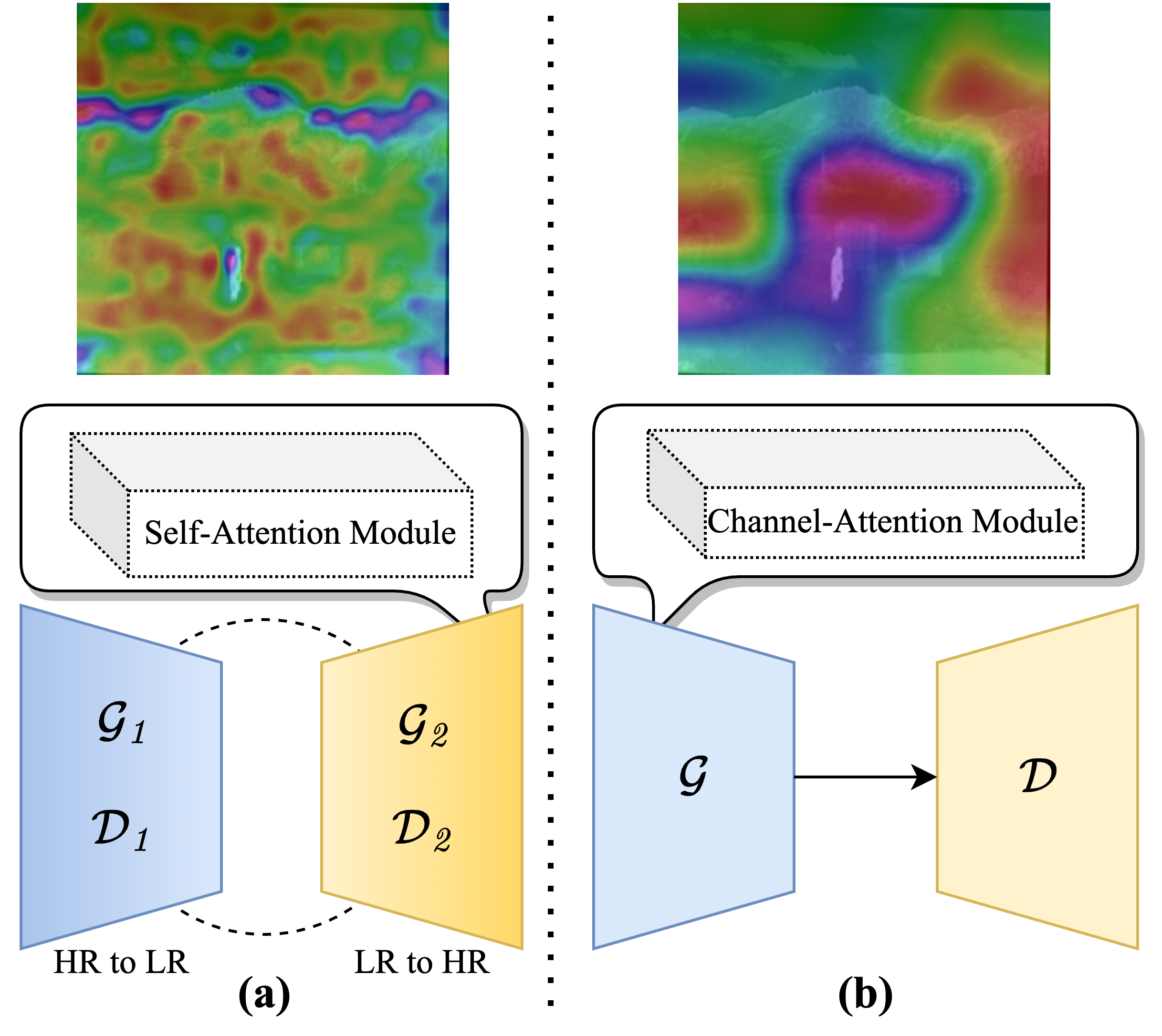}}
\caption{Attentional mechanisms and GAN-based models for IR image super-resolution. Cam Heatmap is also presented.}
\label{fig9}
\end{figure}

Another GAN model based on the attention mechanism has also been proposed, which differs in the loss function and discriminator used\cite{liu2021infrared}. As shown in Fig. \ref{fig9}. Liu, Qing-Ming, et al. use a module based on the channel attention mechanism in the generator at their work. Moreover, they also replaced the original discriminator with a new one designed by the gradient penalty approach from WGAN. It will benefit network convergence. For the loss function, Wasserstein distance is used to evaluate the target image $I_{HR}$ and the reconstructed image $\hat{I}$ (see Eq.\ref{eq.9}).

\begin{equation}
\begin{split}
\mathcal{L_\text{WGAN}} & =E\left[D_\theta\left(I_{H R}\right)-D_\theta\left(G_\theta\left(I_{L R}\right)\right)\right]+, \\
 & \lambda E\left[\left\|\nabla_I D(\hat{I})_2-1\right\|^2\right]
\end{split}
\label{eq.9}
\end{equation}
where the second term on the right side of the equal sign is the gradient penalty term. Furthermore, WGAN is used to optimize the training process is also mentioned in this work\cite{huang2021pricai}.

\textit{2) Introduce extra information:} This approach is focused on the issue that IR images have fewer patterns and further proposes to use visible images to guide the reconstruction. The main options include the following: hybrid and split. For the hybrid model, the features from visible and IR images are not purposely distinguished but mixed directly in the module. In \cite{almasri2018multimodal}, the information is initially extracted by residual blocks and then fused using 1x1 convolution. It is called multimodal, visual-thermal fusion model. Experiments show that visible images are beneficial for improving high-frequency details in IR images. 

For the split model, \textit{Huang et al.\cite{huang2021infrared}} proposed using transfer learning to help introduce the patterns in visible images to reconstruct IR images. In this approach, PSRGAN, two components are proposed: the main path, which is responsible for extracting the features from visible images, and the branch path, which is used to represent the patterns from IR images. Finally, the IR image is reconstructed by a transfer learning strategy that maps both of them to a common feature space. Such an approach can better capture and utilize the features in the model and show better results in the experiments. 

Finally, there are also approaches that focus on the connection between SR methods and other tasks, such as denoising, and then models that combine cross-tasks are proposed. \textit{Batchuluun, Ganbayar, et al.\cite{batchuluun2020deep}} proposed methods that can be used for super-resolution and detection, where the super-resolution component focuses on both denoising and SR. In this work, the number of modules in the generator has been increased. In \cite{mostofa2020joint}, researchers proposed a GAN-based framework for joint cross-modal and super-resolution aerial image vehicle detection. The first sub-network of this approach uses the GAN architecture to generate two different SR images by cross-domain transforms.

In summary, there is no doubt that the GAN model has great advantages in the field of generation, leading to a fairly good performance in SR algorithms. The results achieved in the visible field have attracted the application to the IR image super-resolution. However, according to the experimental results, the direct unmodified introduction of the normal image super-resolution reconstruction method encounters the difficulty of domain shift\cite{shao2018single,guei2018deep,zhang2020single}. The edges in the reconstructed images are not clear. Therefore, new models have been proposed by considering the characteristics of IR images. These models are divided into improved modules\cite{rivadeneira2022novel,liu2021infrared,prajapati2021channel}, introduce extra information\cite{almasri2018multimodal,rivadeneira2020thermal,wu2022infrared}, and multitask\cite{liu2019infrared,batchuluun2020thermal,batchuluun2020deep,mostofa2020joint}. These algorithms have achieved some success, but missing standard benchmarks and datasets that are uniform leads to an unfavorable cross-sectional comparison between different types of methods. In the next section, we present available datasets and evaluation metrics in the IR image super-resolution field.

\begin{table*}[ht]
\centering
\renewcommand\arraystretch{1.1}
\caption{The average results of (PSNR$\uparrow$ MSE$\downarrow$ SSIM$\uparrow$) with scale factor of  4 \& 2 on datasets result-A \& result-C \& CVC10. Best and second-best performances are marked in \textbf{bold} and {\ul underlined}, respectively.}
\resizebox{\textwidth}{!}{%
\begin{tabular}{@{}c|c|c|cccccccccc@{}} % 调整列定义，增加一列
\toprule
\multirow{2}{*}{Scale}       & \multicolumn{1}{c|}{\multirow{2}{*}{Methods}} & \multicolumn{1}{c|}{\multirow{2}{*}{\# Params. (K)}} & \multicolumn{3}{c|}{result-A}                                              & \multicolumn{3}{c|}{result-C}                                              & \multicolumn{3}{c}{CVC10}                             \\ \cmidrule(l){4-12} % 调整横线范围
                             & \multicolumn{1}{c|}{}                         & \multicolumn{1}{c|}{}                         & PSNR$\uparrow $  & MSE$\downarrow $ & \multicolumn{1}{c|}{SSIM$\uparrow $} & PSNR$\uparrow $  & MSE$\downarrow $ & \multicolumn{1}{c|}{SSIM$\uparrow $} & PSNR$\uparrow $  & MSE$\downarrow $ & SSIM$\uparrow $ \\ \midrule
\multirow{19}{*}{$\times 2$} % 调整multirow参数以包含新增行
                             & EDSR\textcolor[RGB]{217,205,144}{\textit{[CVPRW 2017]}}~\cite{lim2017enhanced}                                            & 1,369                                              & 39.0493          & 11.8196          & 0.9414                               & 39.8902          & 8.9865           & 0.9528                               & 44.1770          & 2.7845           & 0.9713          \\
                             & ESRGAN\textcolor[RGB]{217,205,144}{\textit{[ECCVW 2018]}}~\cite{wang2018esrgan}                                        & 16,661                                               & 38.7738          & 12.5212          & 0.9384                               & 39.6111          & 9.5793           & 0.9500                               & 44.0974          & 2.8477           & 0.9709          \\
                             & FSRCNN\textcolor[RGB]{217,205,144}{\textit{[ECCV 2016]}}~\cite{dong2016accelerating}                                        & 475                                               & 39.1175          & 11.3761          & 0.9426                               & 39.9858          & 8.6899           & 0.9535                               & 44.1253          & 2.8162           & 0.9710          \\
                             & SRGAN\textcolor[RGB]{217,205,144}{\textit{[CVPR 2017]}}~\cite{ledig2017photo}                                        & 1,370                                               & 39.0401          & 11.9024          & 0.9414                               & 39.8678          & 9.0586           & 0.9527                               & 44.1736          & 2.7851           & 0.9713          \\
                             & SwinIR\textcolor[RGB]{217,205,144}{\textit{[ICCV 2021]}}~\cite{liang2021swinir}                                        & 11,752                                               & 38.6899          & 12.5694          & 0.9374                               & 39.5215          & 9.6530           & 0.9492                               & 43.9980          & 2.8926           & 0.9704          \\
                             & SRCNN\textcolor[RGB]{217,205,144}{\textit{[T-PAMI 2015]}}~\cite{dong2015image}                                          & 57                                               & 38.9671          & 11.7216          & 0.9414                               & 39.8642          & 8.8857           & 0.9524                               & 44.0038          & 2.9084           & 0.9707          \\
                             & RCAN\textcolor[RGB]{217,205,144}{\textit{[ECCV 2018]}}~\cite{zhang2018image}                                          & 12,467                                               & 38.8145          & 12.4926          & 0.9391                               & 39.7075          & 9.4220           & 0.9511                               & 44.1205          & 2.8170           & 0.9713          \\
                             & PSRGAN\textcolor[RGB]{217,205,144}{\textit{[SPL 2021]}}~\cite{huang2021infrared}                                       & 2,414                                               & 39.2146          & 11.2409          & 0.9429                               & 40.0543          & 8.6101           & 0.9539                               & 44.2377          & 2.7454           & 0.9713          \\
                             & ShuffleMixer(tiny)\textcolor[RGB]{217,205,144}{\textit{[NIPS'22]}}~\cite{sun2022shufflemixer})                            & 108                                               & 39.0465          & 11.7605          & 0.9414                               & 39.8766          & 8.9680           & 0.9527                               & 44.1408          & 2.8113           & 0.9713          \\
                             & ShuffleMixer (base)\textcolor[RGB]{217,205,144}{\textit{[NIPS'22]}}~\cite{sun2022shufflemixer}                            & 121                                               & 38.8066          & 12.3718          & 0.9388                               & 39.6347          & 9.4864           & 0.9503                               & 44.0357          & 2.8809           & 0.9710          \\
                             & HAT\textcolor[RGB]{217,205,144}{\textit{[CVPR 2023]}}~\cite{Chen_2023_CVPR}                                           & 20,624                                               & 38.7754          & 12.4528          & 0.9384                               & 39.6346          & 9.5132           & 0.9500                               & 44.1080          & 2.8244           & 0.9709          \\
                             & RGT\textcolor[RGB]{217,205,144}{\textit{[ICLR 2024]}}~\cite{chen2024recursive}                                              & 10,051                                               & 39.1642          & 11.3382          & 0.9429                               & 40.0522          & 8.6033           & 0.9540                               & 44.2311          & 2.7358           & 0.9717          \\
                             & MambaIR \textcolor[RGB]{217,205,144}{\textit{[ECCV 2024 SOTA]}}~\cite{guo2024mambair}                                              & 20,421                                               & 39.1761          & 11.2081          & \underline{0.9437}                               & 40.1399          & 8.4798           & \underline{0.9544}                              & 44.4181          & 2.6076           & \textbf{0.9720}          \\
                             & ATD\textcolor[RGB]{217,205,144}{\textit{[CVPR 2024]}}~\cite{zhang2024transcending}                                           & 753                                               & 39.0453          & 11.5702          & 0.9432                               & 40.0375          & 8.6155           & 0.9542                              & 44.1901          & 2.7737           & 0.9711          \\
                             & CATANet \textcolor[RGB]{217,205,144}{\textit{[CVPR 2025 SOTA]}}~\cite{liu2025catanet}                                            & 477                                               & 39.0886 & 11.4332 & 0.9430                    & 40.0064 & 8.6387  & 0.9539                      & 44.1656 & 2.7858  & 0.9712 \\
                             % 合并 MambaOut x2
                             & MambaOut\textcolor[RGB]{217,205,144}{\textit{[CVPR 2025 SOTA]}}\cite{yu2024mambaout}                 & 9,669                        & 38.6375          & 12.7091          & 0.9371                               & 39.4900          & 9.7035           & 0.9493                               & 43.9150          & 2.9429           & 0.9704          \\
                             % 合并 VisionMamba x2
                             & VisionMamba\textcolor[RGB]{217,205,144}{\textit{[ICML 2024]}}\cite{zhu2024vision}                   &  27,880                        & 38.7805          & 12.2990          & 0.9392                               & 39.6339          & 9.3781           & 0.9506                               & 43.9521          & 2.9103           & 0.9704          \\
                             % 复制的 IRSRMamba (Ours) 数据行 1
                             & IRSRMamba\textcolor[RGB]{217,205,144}{\textit{[TGRS 2025 SOTA]}}~\cite{11059944}                                           & 26,462                                               & \underline{39.3489} & \underline{10.8767} & \textbf{0.9440}                      & \underline{40.2302} & \underline{8.3164}  & \textbf{0.9548}                      & \underline{44.5310} & \underline{2.5537}  & \textbf{0.9720} \\
                             % 复制的 IRSRMamba (Ours) 数据行 2
                              % 新增的空数据行 for x2
                             &  MambaIRv2 \textcolor[RGB]{217,205,144}{\textit{[CVPR 2025 SOTA]}}~\cite{guo2025mambairv2}                                                                       & 22,903                                               &  39.0421                & 11.5690                  &  0.9423                                    & 39.9295                & 8.7817                 & 0.9531                                    & 44.1442                 & 2.7995                 & 0.9711                 \\
                             & \textbf{GPSMamba} \textcolor[RGB]{217,205,144}{\textit{[2025 SOTA]}}~\cite{huang2025gpsmamba}                                             & 36,942                                               & \textbf{39.3505} & \textbf{10.8608} & \textbf{0.9440}                      & \textbf{40.2418} & \textbf{8.2943}  & \textbf{0.9548}                      & \textbf{44.5543} & \textbf{2.5451}  & \underline{0.9719} \\

                             \midrule
\multirow{19}{*}{$\times 4$} % 调整multirow参数以包含新增行
                             & EDSR\textcolor[RGB]{217,205,144}{\textit{[CVPRW 2017]}}~\cite{lim2017enhanced}                                            & 1,369                                              & 34.5219          & 30.1273          & 0.8548                               & 35.1740          & 23.9917          & 0.8723                              & 40.1190          & 6.8819           & 0.9482          \\
                             & ESRGAN\textcolor[RGB]{217,205,144}{\textit{[ECCVW 2018]}}~\cite{wang2018esrgan}                                         & 16,661                                               & 33.6895          & 34.7337          & 0.8500                               & 34.1650          & 28.9017          & 0.8679                               & 37.9780          & 10.9641          & 0.9455          \\
                             & FSRCNN\textcolor[RGB]{217,205,144}{\textit{[ECCV 2016]}}~\cite{dong2016accelerating}                                        & 475                                               & 33.8556          & 34.4909          & 0.8446                              & 34.5272          & 27.4495          & 0.8636                               & 38.7856          & 9.5482           & 0.9421          \\
                             & SRGAN\textcolor[RGB]{217,205,144}{\textit{[CVPR 2017]}}~\cite{ledig2017photo}                                        & 1,370                                               & 34.5807          & 29.6927         & 0.8556                               & 35.2076          & 23.7701          & 0.8728                               & 40.1479          & 6.8162           & 0.9483          \\
                             & SwinIR\textcolor[RGB]{217,205,144}{\textit{[ICCV 2021]}}~\cite{liang2021swinir}                                        & 11,752                                               & 34.4321          & 30.6081          & 0.8537                               & 35.0329          & 24.6490          & 0.8710                               & 39.9062          & 7.1886           & 0.9479          \\
                             & SRCNN\textcolor[RGB]{217,205,144}{\textit{[T-PAMI 2015]}}~\cite{dong2015image}                                          & 57                                               & 33.6839          & 34.9181          & 0.8415                               & 34.2348          & 28.6115          & 0.8568                               & 38.0976          & 10.7588          & 0.9279          \\
                             & RCAN\textcolor[RGB]{217,205,144}{\textit{[ECCV 2018]}}~\cite{zhang2018image}                                          & 12,467                                               & 34.4280          & 30.8815          & 0.8528                               & 35.0823          & 24.6507          & 0.8705                               & 40.0805          & 6.9225           & 0.9484          \\
                             & PSRGAN\textcolor[RGB]{217,205,144}{\textit{[SPL 2021]}}~\cite{huang2021infrared}                                        & 2,414                                               & 34.4595          & 30.3760          & 0.8540                               & 35.1023          & 24.3147          & 0.8715                               & 39.9533          & 7.1274           & 0.9471          \\
                             & ShuffleMixer(tiny)\textcolor[RGB]{217,205,144}{\textit{[NIPS'22]}}~\cite{sun2022shufflemixer})                            & 108                                               & 34.5440          & 29.9449          & 0.8550                               & 35.1640          & 23.9705          & 0.8723                               & 40.0756          & 6.9296           & 0.9478          \\
                             & ShuffleMixer (base)\textcolor[RGB]{217,205,144}{\textit{[NIPS'22]}}~\cite{sun2022shufflemixer}                            & 121                                               & 34.4507          & 30.6955          & 0.8538                                & 35.0911          & 24.3745          & 0.8714                               & 40.0120          & 7.0622           & 0.9477          \\
                             & HAT\textcolor[RGB]{217,205,144}{\textit{[CVPR 2023]}}~\cite{Chen_2023_CVPR}                                           & 20,624                                               & 34.4947          & 30.4086          & 0.8542                               & 35.1239          & 24.3103          & 0.8713                               & 40.0934          & 6.9078           & 0.9478          \\
                & RGT\textcolor[RGB]{217,205,144}{\textit{[ICLR 2024]}}~\cite{chen2024recursive}                                             & 10,051                                               & 34.3826          & 31.0046          & 0.8535          & 35.0534          & 24.5924          & 0.8711                & 39.8420          & 7.3060           & 0.9472           \\
                            & MambaIR\textcolor[RGB]{217,205,144}{\textit{[ECCV 2024 SOTA]}}~\cite{guo2024mambair}                                           & 20,421                                               & 34.0267          & 32.9760          & 0.8510                               & 34.5662          & 27.0850          & 0.8681                               & 38.1878          & 10.8653          & 0.9404          \\
                              & ATD\textcolor[RGB]{217,205,144}{\textit{[CVPR 2024]}}~\cite{zhang2024transcending}                                           & 753                                               & 34.6113          & 29.5294          & 0.8569                              & 35.2347          & 23.6882          & 0.8737                              & 40.2897          & 6.6001           & 0.9494      \\
                            & CATANet \textcolor[RGB]{217,205,144}{\textit{[CVPR 2025 SOTA]}}~\cite{liu2025catanet}                                           & 477                                               & 34.0613 & 32.7100  & 0.8447                    & 34.8947 & 25.3673  & 0.8663                      & 39.0302 & 8.8215  & 0.9405 \\
                            % 合并 MambaOut x4
                             & MambaOut\textcolor[RGB]{217,205,144}{\textit{[CVPR 2025 SOTA]}}\cite{yu2024mambaout}                 & 9,669                        & 34.4483          & 30.6792          & 0.8527                               & 35.0456          & 24.7055          & 0.8698                               & 40.1587          & 6.8076           & 0.9485          \\
                             % 合并 VisionMamba x4
                             & VisionMamba\textcolor[RGB]{217,205,144}{\textit{[ICML 2024]}}\cite{zhu2024vision}                   & 27,880                        & 34.5941          & 29.5650          & 0.8564                               & 35.2327          & 23.6467          & 0.8733                               & 40.1705          & 6.7933           & 0.9484          \\
                             & IRSRMamba\textcolor[RGB]{217,205,144}{\textit{[TGRS 2025 SOTA]}}~\cite{11059944}                                     & 26,462                                               & \underline{34.6755} & \underline{29.0551} & \underline{0.8577}                      & \underline{35.3074} & \underline{23.2857} & \underline{0.8745}                      & \underline{40.4052} & \underline{6.4536}  & \underline{0.9497} \\
                              % 新增的空数据行 for x4
                             &  MambaIRv2 \textcolor[RGB]{217,205,144}{\textit{[CVPR 2025 SOTA]}}~\cite{guo2025mambairv2}                                                                       & 22,903                                               &  29.5295                &  97.3011                &  0.8420                                    &  28.6532                & 130.8167                  & 0.8577                                     & 27.2805                 &  121.7986                &  0.9315               \\
                             % 复制的 IRSRMamba (Ours) 数据行 1
                             & \textbf{GPSMamba} \textcolor[RGB]{217,205,144}{\textit{[2025 SOTA]}}~\cite{huang2025gpsmamba}                                             & 36,942                                               & \textbf{34.7421} & \textbf{28.7268} & \textbf{0.8587}                      & \textbf{35.4007} & \textbf{22.9636} & \textbf{0.8756}                      & \textbf{40.5475} & \textbf{6.2681}  & \textbf{0.9503} \\

                             \bottomrule
\end{tabular}%
}
\label{tab.2}
\end{table*}

\begin{table*}[!t]
\centering
\caption{Systematic comparison of common datasets for Infrared Image Super-Resolution. The table contrasts datasets on key attributes including the number of images, resolution, modality type (IR-only vs. IR-VIS pairs), and primary application domain.}
\label{tab_dataset_comparison}
\renewcommand\arraystretch{1.4} % Adjust row height for readability
\small % Use a slightly smaller font to ensure the table fits well
\begin{tabular}{@{}l l c c l p{5cm}@{}}
\toprule
\textbf{Dataset Name} & \textbf{Reference} & \textbf{\# Images} & \textbf{Resolution} & \textbf{Modality} & \textbf{Application Domain / Notes} \\
\midrule
\rowcolor[HTML]{EFEFEF} 
NATO SET-140 & \cite{weiss2012standard} & 140 sequences & \begin{tabular}[c]{@{}c@{}}320$\times$256 \\ 640$\times$480\end{tabular} & IR-only & \textbf{Surveillance/Military}: Contains various target/background settings with object and camera movement. \\

CASIA Interval v3 & \cite{alonso2017iris} & 2,655 & 280$\times$320 & IR-only (NIR) & \textbf{Biometrics}: Near-infrared iris images from 249 subjects. LR pairs are generated via downsampling. \\

\rowcolor[HTML]{EFEFEF} 
ULB17-VT & \cite{almasri2018multimodal} & 570 pairs & 240$\times$320 & IR-VIS Pairs & \textbf{General/Surveillance}: Manually extracted and annotated thermal and visible image pairs. \\

IR-COLOR2000 & \cite{du2020rgb} & 2,000 pairs & 128$\times$128 & IR-VIS Pairs & \textbf{General}: A large-scale dataset of paired infrared and color images for training. \\

\rowcolor[HTML]{EFEFEF} 
IR100 & \cite{huang2021infrared} & 100 & 640$\times$480 & IR-only & \textbf{Surveillance (Pedestrian)}: Sourced from the FIR Sequence Pedestrian Dataset. LR pairs are generated. \\

CVC-09-1K & \cite{huang2021pricai} & 1,000 & 640$\times$480 & IR-only & \textbf{Autonomous Driving}: Randomly selected from the FIR Dataset. LR pairs are generated. \\

\rowcolor[HTML]{EFEFEF} 
Rivadeneira et al. & \cite{rivadeneira2019thermal} & Not Specified & 640$\times$512 & IR-only & \textbf{General/Outdoor}: Acquired with a TAU2 thermal camera. LR pairs must be generated. \\

CDN-MRF & \cite{he2018cascaded} & Not Specified & 640$\times$480 & IR-only & \textbf{General/Surveillance}: Covers a wide range of scenarios (vehicles, pedestrians, buildings). LR pairs are from the real world. \\
\bottomrule
\end{tabular}
\label{tab.dataset1}
\end{table*}

\begin{table}[!t]
\centering
\caption{\rr{Condensed summary of standard benchmarking protocols for popular IRSR datasets. The table outlines common practices to facilitate fair comparison and reproducibility.}}
\label{tab_benchmark_protocols_condensed}
\renewcommand\arraystretch{1.3}
% Use resizebox to scale the table to the width of a single column
\resizebox{\columnwidth}{!}{%
\begin{tabular}{@{}llll@{}}
\toprule
\textbf{Dataset} & \textbf{Train/Test Split} & \textbf{Downsampling} & \textbf{Preprocessing} \\
\midrule
\rowcolor[HTML]{EFEFEF} 
\textbf{FLIR ADAS} & Train: 8,862 / Test: 1,364 & Bicubic ($\times$2, $\times$3, $\times$4) & Use grayscale thermal; crop watermarks; normalize. \\

\textbf{KAIST} & Official train/test split & Bicubic ($\times$2, $\times$3, $\times$4) & Extract thermal channel; normalize. \\

\rowcolor[HTML]{EFEFEF} 
\textbf{CVC-09-1K} & Training set only & Bicubic ($\times$2, $\times$4) & Evaluate on original HR from downsampled input. \\
\bottomrule
\end{tabular}%
}
\label{tab.dataset2}
\end{table}

\subsubsection{Transformer-based}

\tcmi{Transformer-based methods have transformed IRSR by addressing challenges unique to IR imaging, such as low contrast and limited gradients. Leveraging self-attention mechanisms\cite{khan2022transformers}, these models excel in capturing long-range dependencies, making them ideal for reconstructing high-resolution IR images with superior structural consistency.}

\rr{As illustrated in Fig. \ref{fig8} (b), Transformers overcome the inherent locality bias of CNNs, whose fixed, local receptive fields constrain long-range dependency modeling. Architectures like SwinIR employ a windowed self-attention mechanism for computational efficiency. Crucially, they introduce a shifted-window strategy that alternates window configurations across layers. This enables cross-window connections, efficiently building a global receptive field without the quadratic complexity of standard Vision Transformers. This explicit and efficient global modeling is particularly effective for IRSR, ensuring the structural consistency of large, homogeneous thermal regions.}

\tcmi{The architectural innovations within Transformer-based IR SR models demonstrate a strong focus on achieving efficient global feature representation. Swin Transformer-based models, such as SwinIBSR\cite{shi2024swinibsr} and SwinIPISR\cite{wu2023swinipisr}, extend the Swin architecture with residual connections and tailored degradation models, improving generalization and robustness in real-world scenarios. Similarly, LKFormer replaces conventional self-attention with large kernel convolutional attention blocks, striking a balance between computational efficiency and non-local feature modeling\cite{qin2024lkformer}. These advancements reflect a shared objective to optimize performance while maintaining scalability.}

\tcmi{A defining characteristic of these methods is their adaptability to specific tasks and modalities. Techniques like Cross-modal Texture Transformers (CMTT) effectively transfer information from visible-spectrum images to improve IR image quality, while edge-focused models such as TESR incorporate auxiliary networks to enhance texture and detail recovery\cite{jiang2024cross}. Meanwhile, DASR utilizes dual-attention mechanisms to simultaneously capture local and global information, underscoring the versatility of Transformer architectures in addressing diverse challenges\cite{liang2023dasr}.}

\tcmi{Compared to GANs, which excel in texture synthesis but are prone to artifacts under domain shifts, Transformer-based methods demonstrate superior robustness through explicit global contextual modeling. Similarly, while CNNs are computationally efficient and effective in learning localized features, Transformers outperform in scenarios requiring long-range dependency modeling and structural consistency. These distinctions make Transformers particularly suitable for dynamic environments where IR imaging data is sparse or noisy. For instance, TnTViT-G\cite{mehri2023tntvit} demonstrates how visible-spectrum guidance can compensate for missing IR image details, exemplifying the potential of cross-spectral applications.}

\tcmi{In summary, Transformer-based methods represent a significant advancement in IRSR by combining efficient feature extraction with innovative attention mechanisms. However, their computational complexity remains a challenge, particularly for high-resolution IR image reconstruction, where scaling attention mechanisms can be resource-intensive. Addressing these limitations could further enhance their applicability in real-time systems. These approaches share a unified goal of overcoming the limitations of IR imaging systems while optimizing for scalability and practical deployment. Future research will likely explore deeper integration with emerging imaging modalities and further enhancements for real-time applications.}

% While Transformer-based models excel at capturing global context, their self-attention mechanism has a computational complexity that is quadratic with respect to the number of image patches ($O(N^2)$). This can become a bottleneck for high-resolution image processing. This limitation has spurred research into more efficient architectures. State-space models (SSMs), such as Mamba\cite{huang2024irsrmamba}, have recently emerged as a powerful alternative. SSMs offer linear-time complexity ($O(N)$) for sequence modeling, making them highly scalable. When adapted for vision tasks, they promise to balance long-range dependency modeling with computational efficiency. For IRSR, the trade-off is clear: Transformers offer a well-established and powerful, albeit computationally expensive, method for global feature integration, whereas SSMs represent a more nascent but potentially more scalable path for future high-resolution and real-time applications. A key open question is how effectively the 1D-sequence processing nature of SSMs can be adapted to capture the complex 2D spatial relationships inherent in images compared to the explicit 2D windowing schemes in vision transformers. 

\rr{The quadratic computational complexity ($O(N^2)$) of the self-attention mechanism in Transformers, while effective for global context modeling, presents a bottleneck for high-resolution image processing\cite{huang2024irsrmamba}. This limitation has spurred research into more efficient architectures, with State-space models (SSMs) like Mamba emerging as a powerful alternative. SSMs offer linear-time complexity ($O(N)$) for sequence modeling, balancing long-range dependency capture with high computational efficiency. However, a key research challenge remains in adapting the inherent 1D sequential processing of SSMs to effectively model complex 2D spatial relationships—a task for which vision transformers' explicit 2D windowing schemes are well-established. Therefore, while Transformers are a robust solution, SSMs represent a nascent but potentially more scalable paradigm for future high-resolution and real-time IRSR applications.}

\subsubsection{Diffusion model-based}

\rr{Diffusion models offer a powerful generative approach for IRSR. By learning to reverse a stochastic noising process, they excel at modeling the complex degradation and noise patterns inherent to thermal imagery. State-of-the-art methods, such as DifIISR\cite{li2025difiisr}, demonstrate that tailoring the reverse diffusion process with task-specific guidance—for instance, by incorporating priors on the infrared thermal spectrum distribution—can significantly enhance reconstruction fidelity. Although computationally demanding, their ability to generate high-fidelity results by learning the underlying data distribution positions them as a leading approach in the field.}

\rr{Beyond reconstruction fidelity, the practical deployment feasibility of IRSR models is critically dependent on their computational complexity. To provide a clear overview of the trade-offs between performance and efficiency, we present a comparison of representative learning-based models in terms of their parameter counts (\# Params.). Table.\ref{tab.2}, adapted from the recent work by \textit{Huang et al}\cite{huang2025gpsmamba}, summarizes these metrics alongside performance on benchmark datasets. The analysis in Table.\ref{tab.2} reveals a clear trend: top-performing models such as IRSRMamba (26.5M), and GPSMamba (36.9M) carry a substantial number of parameters. This high complexity is often necessary to achieve state-of-the-art fidelity. In contrast, lightweight models like FSRCNN (475K) and ShuffleMixer (tiny) (108K) offer significantly greater efficiency, making them suitable for resource-constrained environments, albeit with a compromise in reconstruction quality. This comparison provides a practical guide for researchers and practitioners to select an appropriate model that balances their specific needs for performance against the available computational budget.}

In summary, the evolution of IRSR methods reflects a clear trajectory from model-driven to data-driven approaches. Traditional methods, while interpretable, struggled with the complex and non-linear nature of IR image degradation. The advent of deep learning, particularly CNNs, marked a paradigm shift, enabling the automatic learning of intricate mappings from LR to HR. GANs further advanced the field by improving perceptual quality and texture synthesis, which is critical for low-contrast thermal scenes. Most recently, Transformer and state-space architectures are pushing the boundaries by effectively modeling long-range dependencies, leading to improved structural integrity in reconstructed images. This progression demonstrates a continuous effort to design models that are not only powerful but also increasingly tailored to the unique statistical properties and challenges of infrared imagery.

\section{Datasets \& Metrics}

We will present the datasets and image quality assessment metrics applied in the IR image super-resolution, respectively. 

\subsection{Datasets}
For the super-resolution study, the datasets included: the training dataset and test dataset.  Considering the training dataset, we hope that the samples are large enough to cover the current natural and realistic situations. It will be beneficial to the models' generalization. For the test dataset, we would like to have standard datasets corresponding to different tasks. For example, in the face reconstruction task, the face dataset is important. But for iris reconstruction, the iris dataset would be more useful.

% Datasets for IR images and video are available in other research topics, but fewer are focused on super-resolution. 

\tcmi{Datasets for IR images and video are available in other research topics\cite{danaci2022survey}, but fewer are focused on super-resolution. Despite the availability of datasets such as IR100 and NATO SET-140, current datasets often lack paired samples that span diverse environmental conditions or integrate multi-modal information. Future efforts should prioritize creating datasets reflecting real-world challenges, such as adverse weather conditions, varying lighting, and diverse environmental scenarios. Additionally, generative models offer a promising solution to data scarcity by synthesizing diverse training samples, and enhancing model robustness.} We will present the following representative work: \textit{Weiss et al.\cite{weiss2012standard}} present a publicly distributable dataset for the assessment of fusion, super-resolution, local contrast enhancement, dynamic range compression, and performance of image-based non-uniformity correction (NUC) algorithms. This dataset records images for different target background settings, camera or object movements, and temperature contrasts. Furthermore, \textit{Alonso-Fernandez et al.} present 1,872 NIR iris image datasets\cite{alonso2015eigen}. \textit{Almasi \& Debeir\cite{Almasri2018MultimodalSF}}, on the other hand, presented a benchmark ULB17-VT dataset containing thermal images and their corresponding visual images. \textit{Du, Juan, et al.\cite{du2020rgb}} also presented their dataset. For small sample reconstruction, \textit{Huang, Yongsong, et al.} proposed the IR100 dataset\cite{huang2021infrared}. \textit{Huang, Yongsong, et al.\cite{huang2021pricai}} further proposed a universal dataset including 1000 samples. The details are shown in Tab.\ref{tab.dataset1}. \rr{Moreover, Table.\ref{tab.dataset2} outlines the widely adopted training and testing splits for prominent datasets like FLIR ADAS and KAIST, details the standard downsampling method (bicubic) and typical scale factors used to generate low-resolution inputs, and describes common preprocessing steps such as channel selection and value normalization.}

Despite these efforts, significant weaknesses persist in existing datasets. These include: \textbf{1) Lack of Standardization:} Datasets often vary widely in sample size and fail to document crucial acquisition parameters (e.g., sensor type, lens specifications), making direct, fair comparisons between methods challenging. \textbf{2) Scarcity of Paired Multi-Modal Data:} Few datasets provide perfectly aligned high-resolution visible image pairs, hindering research on multi-modal fusion. \textbf{3) Over-reliance on Synthetic Degradation:} Crucially, the majority of these benchmarks create LR-HR pairs via simple synthetic degradations like bicubic downsampling. This approach completely fails to capture the complex, modality-specific artifacts found in real-world IR systems, such as non-uniformity noise, dead pixels, and atmospheric effects. Establishing a benchmark with authentic, hardware-induced degradations would be a monumental contribution to the field, enabling more rigorous and meaningful algorithmic evaluation.

For the test dataset, it is very common to select a small sample from the training dataset. These images are used as reference samples to evaluate the algorithm performance. On the other hand, IR image super-resolution is to be used in fused images sometimes. Considering this situation, there are also some algorithms\cite{huang2021infrared,huang2021pricai} that use fused images as reference, such as result-A\cite{liu2018infrared1} and result-C\cite{zhang2017infrared1}. Different types of IR image samples will be useful to evaluate the reconstruction algorithm's robustness.

There are still some weaknesses in these datasets, which include the following aspects: these data are collected without indicating the relevant equipment parameters, which could influence the dataset selection in real situations. In addition, these datasets have different sample sizes, making direct comparisons between methods challenging. Furthermore, some datasets lack matching high-resolution visible image pairs, hindering research on multi-modal fusion. Crucially, the majority of these benchmarks rely on synthetic degradation (e.g., bicubic downsampling) to create LR-HR pairs. This approach fails to capture the complex, modality-specific degradations found in real-world scenarios. Establishing a benchmark dataset containing authentic sensor noise, optical artefacts, and non-uniformity effects from various hardware would be a significant contribution, facilitating more rigorous and meaningful algorithmic evaluation. For datasets, we can make more efforts to complete the standard. On the other hand, it is also important to evaluate the reconstructed image quality. Then researchers also started to focus on proposing evaluation metrics applicable to infrared images which are also called IQA metrics.

\subsection{Metrics} 

In the field of image reconstruction, how to evaluate reconstructed images is also a popular topic. We usually have three types of image evaluation metrics: reference metrics, non-reference metrics, and human evaluation.

\subsubsection{Reference metrics} 

For reference metrics, the most commonly used to evaluate the algorithms' performance are peak signal-to-noise ratio (PSNR) and structural similarity index (SSIM). First, PSNR uses the maximum pixel value (denoted as $L$) and the mean square error (MSE) between images to estimate the reconstructed image quality. As shown in Eq. \ref{eq.13}:

\begin{equation}
\mathrm{PSNR}=10 \cdot \log _{10}\left(\frac{L^2}{\frac{1}{N} \sum_{i=1}^N(I(i)-\hat{I}(i))^2}\right)
\label{eq.13}
\end{equation} where $L$ equals to 255 in general cases using 8-bit representations, $I$ denotes the high resolution image while $\hat{I}$ represents the reconstructed image. Since PSNR is only related to pixel-level MSE and only cares about the differences between corresponding pixels without concerns about visual perception, it tends to lead to poor performance in reconstruction quality in real scenes, while we are usually more concerned with human perception. But in order to compare with other work, we always need to calculate this metric. Then, the metric closer to the human subjective evaluation was proposed: SSIM\cite{wang2004image}.

SSIM is dedicated to measuring structural similarity between images, based on independent comparisons with brightness $\mathcal{C}_l(I, \hat{I})$, contrast $\mathcal{C}_c(I, \hat{I})$ and structural $\mathcal{C}_s(I, \hat{I})$ aspects (see Eq.\ref{eq.14}). 

\begin{equation}
\operatorname{SSIM}(I, \hat{I})=\left[\mathcal{C}_l(I, \hat{I})\right]^\alpha\left[\mathcal{C}_c(I, \hat{I})\right]^\beta\left[\mathcal{C}_s(I, \hat{I})\right]^\gamma
\label{eq.14}
\end{equation} where $\alpha, \beta, \gamma$ are control parameters for adjusting the relative importance. More calculation details can be found in \cite{wang2004image}. In short, SSIM evaluates the reconstruction quality from the perspective of human visual system\cite{wang2002image}. Therefore, it better meets the requirements for perceptual evaluation\cite{sheikh2006statistical,wang2009mean} and has been widely used in IR image super-resolution. In summary, the assessment of reference metrics always requires ground-truth images for comparison. In the real world, there is no ground-truth image. Then, no reference metrics are proposed.

\subsubsection{No reference metrics}

We will briefly describe the representative work of no reference metrics. The Natural Image Quality Evaluator (NIQE)\cite{mittal2012making} is a statistical collection of "quality-aware" features built on a simple but successful Natural Scene Statistics (NSS) model in the spatial domain. The researchers also evaluated the perceptual image patch similarity (LPIPS) directly through the trained deep network based on the difference in depth features. According to the experimental results, it would be fairer to use the features extracted by the CNN to evaluate the image quality\cite{zhang2018unreasonable}.

\subsubsection{Human evaluation}

Human subjective evaluation is the best indicator. In general, we need to evaluate and then score the reconstructed images. However, in order to exclude errors due to visual fatigue, we need more participants and more time to remove biased data. In other words, the time and resource cost for this evaluation method would be the most expensive.

\subsubsection{Assessment of IR image super-resolution}

For IR images, which suffer from missing rich details and colors, directly applying metrics designed for visible images can be misleading. While no-reference metrics like NIQE and LPIPS avoid the need for a ground-truth, they are typically trained on natural scene statistics of visible images and may not align with the perceptual qualities of thermal imagery. The development of IR-specific metrics is therefore crucial. For instance, in \cite{sun2022infrared}, researchers proposed a hybrid metric $Q_{SR}$... However, the field still lacks a widely accepted no-reference metric designed explicitly to quantify key thermal attributes. An ideal metric would be sensitive to thermal contrast enhancement, the suppression of characteristic noise patterns (e.g., stripe noise), and the preservation of fine thermal gradients, which are often more critical for downstream tasks than overall "naturalness".

\begin{equation}
Q_{\mathrm{SR}}=\lambda_1 \frac{1}{\log _{10}\left(S_{\mathrm{BRISQUE}}\right)}+\lambda_2 S_{\mathrm{PSNR}}
\label{eq.15}
\end{equation}

\section{Future Trends and Discussion}

In this work, we systematically review the excellent previous work on IR image super-resolution and make a comparison with the summary. Thus, we will introduce some attractive directions in this field and ideas that will further improve the model performance in this section.

\subsection{New models \& new paradigms.}

Good model design can improve the reconstructed image quality by better nonlinear mapping to represent complex patterns in high-dimensional space. \hyss{As we embrace the advent of Large Language Models (LLMs)  in the realm of natural language processing, it is also pertinent to consider the potential impact that the forthcoming Large Vision Models (LVMs) may have on the task of IR image super-resolution (IRSR).} For the new paradigm, we expect that the correlation between low and high frequency information in IR images will be studied to explain the correlation between them. We will present more details below.

\tcmi{\textit{Generative Models.} \hyss{With the rapid advancement of deep learning, state-space models and diffusion models\cite{rombach2022high,salimans2022progressive,lyu2022accelerating,cao2022survey} are gaining increasing attention in the community. State-space models, known for their efficient handling of long-range dependencies and scalability, offer a compelling alternative to traditional Transformer architectures for low-level vision tasks. We anticipate that the classical challenge of IR image super-resolution will significantly benefit from the integration of these state-of-the-art generative models.}}

\tcmi{\textit{Data Augmentation.} A novel data augmentation approach for guided thermal image super-resolution generates synthetic thermal-like images from visible spectrum inputs using CycleGAN, effectively aligning the guiding and guided domains to enhance model performance\cite{suarez2024enhancement}. Looking ahead, the development of advanced data augmentation paradigms, including domain-specific synthetic data generation and cross-modality transformations, represents a promising direction for overcoming data scarcity and improving model robustness in real-world applications.}

\textit{Large Vision Models.} Large Language Models have revolutionized the field of natural language processing, enabling machines to understand and generate human-like text. Drawing parallels, the development of Large Vision Models can have a profound impact on the field of computer vision, particularly in tasks like IR image super-resolution. This can mitigate the requirement for substantial amounts of infrared-specific training data, which is often challenging to procure. However, a significant challenge lies in adapting these massive models, typically pretrained on billions of visible-light images, to the comparatively small and domain-specific IRSR datasets without catastrophic forgetting or overfitting. To bridge this gap, several concrete strategies are emerging as promising research directions:

\begin{itemize}
    \item \textbf{Parameter-Efficient Fine-Tuning:} Techniques like Low-Rank Adaptation (LoRA)\cite{hu2022lora} allow for the fine-tuning of LVMs by updating only a small fraction of their parameters. This drastically reduces computational costs and memory requirements, making it feasible to adapt LVMs on modest-sized IR datasets.
    \item \textbf{Domain-Adaptive Pretraining:} Instead of direct fine-tuning, a two-stage approach can be adopted. First, the LVM is pretrained on a large corpus of unlabeled IR images (if available) to adapt its features to the thermal domain. Subsequently, it is fine-tuned on the smaller, paired IRSR dataset.
    \item \textbf{Feature Distillation:} A smaller, task-specific IRSR network can be trained to mimic the feature representations or output of a powerful LVM. This allows the smaller model to inherit the rich prior knowledge of the LVM without inheriting its massive parameter count, making it suitable for deployment.
    \item \textbf{Others:} \rr{Beyond the adaptation of large-scale models, several other emerging paradigms are set to drive future progress in IRSR. First, generative diffusion models offer a powerful solution for synthesizing plausible high-frequency textures, directly addressing the tendency of conventional methods to produce overly smooth results. Future work should focus on adapting the generative process to the unique statistical properties of infrared imagery through domain-specific conditioning. Second, cross-modal pretraining, inspired by models like CLIP\cite{radford2021learning}, can provide strong semantic priors to guide reconstruction, ensuring that super-resolved outputs are not only visually sharp but also contextually consistent.}
\end{itemize}

\rr{\textit{Multi-Modal Fusion.} Multi-sensor fusion is an emerging strategy to regularize the ill-posed problem of infrared super-resolution. By conditioning the reconstruction process on auxiliary data, these methods leverage strong priors to compensate for the high-frequency information deficit inherent to single-modality IR imagery. A primary approach fuses IR with co-registered visible-light imagery. The visible-light stream provides rich textural and structural priors, which are integrated into the IR upsampling network via mechanisms such as cross-attention. This allows for the synthesis of fine-grained details that are otherwise non-recoverable from the low-resolution thermal data, a crucial capability for applications requiring high-fidelity scene understanding like autonomous driving. Alternatively, fusing thermal data with multispectral or hyperspectral information is pivotal for applications demanding spectral fidelity, such as remote sensing. The outcome is a product that is not only spatially enhanced but also spectrally accurate, enabling reliable quantitative analysis.}

Investigating these strategies will be key to unlocking the full potential of LVMs for IRSR and enhancing the practical relevance of this promising outlook.

\subsection{New application} 

With the development in the field of image super-resolution, blind super-resolution is the future-oriented direction\cite{zhang2022closer,li2022face,yang2022degradation,zhou2022joint}. Although many models or algorithms have been proposed for image super-resolution, the performance in the real world is not ideal. The main problem is that these algorithms simply summarize the degradation as bicubic downsampling and ignore the blurred kernels that exist in the real world, such as noise and compression\cite{lee2022learning,huang2022rethinking,zhang2021designing,wang2021real,son2021toward}. Thus, studies on real-world super-resolution of visible images began to attract the attention of researchers. In IR images, to make the algorithms applicable in the real world, we can pay more attention to the degradation models of IR imaging and design the corresponding reconstruction algorithms. All these works will help the algorithms to face the complex degradation environment. A key enabler for this will be the development of more sophisticated, physically-grounded degradation models that go beyond simple bicubic downsampling to accurately simulate sensor noise, atmospheric turbulence, and thermal drift. Such models are crucial for training robust blind SR networks.
\textbf{Explainable AI (XAI) for IRSR.} As deep learning models become more complex (e.g., LVMs), their "black-box" nature becomes a significant concern, especially in critical applications like medical diagnosis or autonomous navigation. Future research should explore XAI techniques to understand which image features an IRSR model relies on to reconstruct details. This can help diagnose model failures, build trust, and potentially guide the design of more robust and reliable networks.
\textbf{Energy-Efficient and On-Device IRSR.} \rr{Many practical applications of IR imaging, such as in drones, portable thermal cameras, or edge devices in vehicles, are resource-constrained. This creates a strong demand for lightweight and energy-efficient SR algorithms. Future work should focus on model compression, quantization, and designing efficient network architectures that can perform high-quality super-resolution in real-time on devices with limited computational power and memory.}

\subsection{Dataset and evaluation metrics} 

On the one hand, datasets, the cornerstone for algorithm training, are fundamental to this task. As mentioned before, the lack of standardized, comprehensive datasets remains a significant barrier. Beyond simply increasing sample size, future efforts must focus on curating benchmarks that reflect real-world conditions. This involves not only documenting optical device details and capture parameters but, more importantly, capturing paired LR-HR images from actual hardware under varying operational conditions. Such a dataset would contain authentic degradations, including fixed-pattern noise (FPN), dead pixels, thermal drift, and atmospheric attenuation, providing a far more realistic testbed for blind and real-world SR algorithms than current synthetically degraded datasets. It will be beneficial to compare the performance between different types of algorithms. If a matching dataset with visible images, captured under the same environmental conditions, could also be included, there is no doubt that it would help to validate the effectiveness of introducing visible image patterns. Moreover, available datasets are also provided for thermal infrared image challenges as well. For example, the Thermal Image
Super-Resolution (TISR) challenge organized in the Perception Beyond the Visible Spectrum (PBVS) is one of the promising events. This challenge provides the training dataset and the test dataset with 951 and 50 samples, respectively\cite{rivadeneira2022thermal,wang2022cippsrnet}.

On the other hand, developing reliable evaluation metrics remains a critical challenge. Future work must move beyond PSNR and SSIM, which are poorly correlated with both human perception of thermal imagery and performance in task-oriented applications (e.g., detection, tracking). The community needs to articulate and adopt no-reference metrics that are specifically sensitive to infrared contrast, noise statistics, and edge fidelity in low-gradient regions. Such metrics would enable more meaningful algorithm comparisons and guide the development of models that generate genuinely useful high-resolution thermal images, rather than those that simply optimize for pixel-wise similarity.

\rr{Specifically, this new section synthesizes our review by charting a course for future research, organized around three interconnected pillars. We begin by examining the architectural frontier, exploring the transformative potential of adapting Large Vision Models to the infrared domain through parameter-efficient techniques like LoRA. This discussion extends to the promise of generative models, such as state-space and diffusion architectures, for synthesizing high-fidelity textures, and considers the continued importance of multi-modal fusion. These advanced models, however, must be directed toward solving critical, real-world challenges, propelling the field beyond idealized assumptions toward robust blind IRSR. Concurrently, we address the growing necessity for Explainable AI (XAI) to foster trust in critical systems, alongside the practical demand for energy-efficient algorithms suitable for on-device deployment. Underpinning all such progress is the need to address foundational limitations. Therefore, we conclude by advocating for two critical shifts: the curation of realistic datasets that capture authentic hardware degradations, and the development of new evaluation metrics that better correlate with human perception and task-specific performance, moving beyond the confines of PSNR and SSIM.}

\section{Conclusion}

In this paper, we provide a comprehensive survey of IR image super-resolution research from the past two decades. We discuss its fundamental role in engineering applications and analyze key factors limiting IR imaging quality, such as noise and hardware constraints, highlighting the high costs associated with hardware redesign. We systematically classify and summarize both traditional algorithms and deep learning-based methods, alongside essential datasets and image quality assessment metrics. Our review bridges the gap between traditional super-resolution approaches and the unique challenges of IR imaging, offering actionable insights for developing domain-specific algorithms. Future research could explore integrating Large Vision Models and advanced data augmentation techniques to address current limitations and expand the applicability of IR image super-resolution in diverse real-world scenarios.

% Finally, we point out some existing challenges for the future. We hope this survey will not only provide a better understanding of IR image super-resolution but also contribute to future development in this field and the real-world application of related algorithms. 

\section*{Acknowledgments}
This work was supported by JSPS KAKENHI Grant Numbers JP23KJ0118, JP23K11176, and JP25K03130. All data included in this study are available upon request by contact with the corresponding author.

%\balance
\section*{}
\bibliographystyle{ieeetr} 
\bibliography{IEEEabrv,ref}

%%%%%%%%%
\begin{IEEEbiography}[{\includegraphics[width=1in,height=1.25in,clip,keepaspectratio]{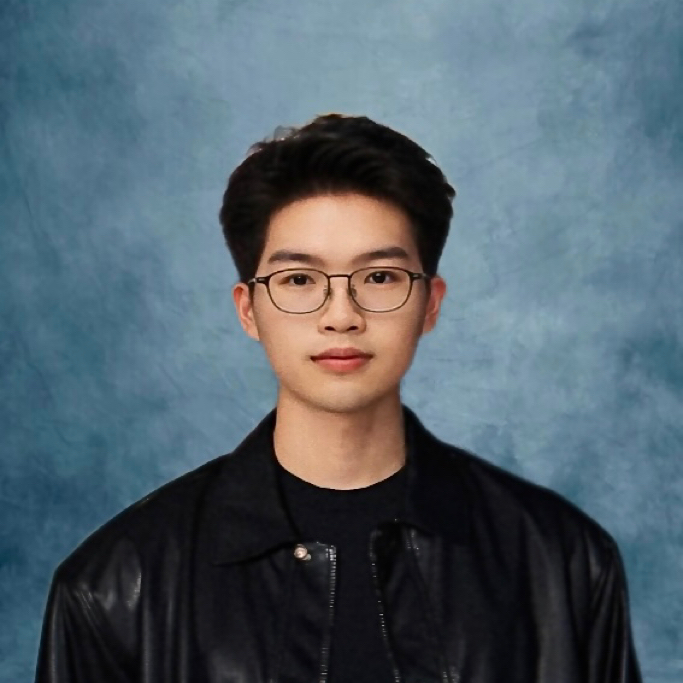}}]{Yongsong Huang}(Member, IEEE) received the B.E. and M.E. degrees from Guilin University of Electronic Technology, Guilin, China, in 2018 and 2021, respectively, and the Ph.D. degree from Tohoku University, Sendai, Japan, in 2025.
He was a JSPS Research Fellow and has held visiting researcher positions at Harvard University, Cambridge, MA, USA; Massachusetts General Hospital, Boston, MA, USA; and National Taiwan University, Taipei City, Taiwan. He is an Assistant Professor at the Advanced Institute of Convergence Knowledge (So-Go-Chi) Informatics, Tohoku University. He is a co-author of the book Applications of Generative AI (Springer). His research interests include computer vision, data science, and pattern recognition. Dr. Huang was awarded the JSPS Special Research Grant Allowance in 2024.
\end{IEEEbiography}

\begin{IEEEbiography}[{\includegraphics[width=1in,height=1.25in,clip,keepaspectratio]{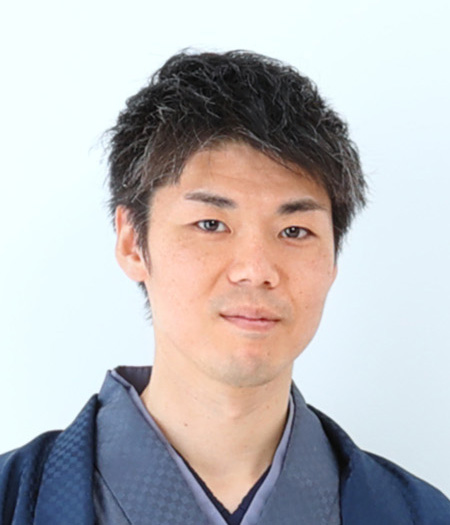}}]{Tomo Miyazaki} 
(Member, IEEE) received B.E. and Ph.D. degrees from Yamagata University and Tohoku University in 2006 and 2011, respectively. He worked on a Geographic Information System at Hitachi, Ltd until 2013. He was a postdoctoral researcher and an assistant professor from 2013 to 2023 at Tohoku University. Since 2024, he has been an Associate Professor. His research interests include pattern recognition and image processing.
\end{IEEEbiography}

% \begin{IEEEbiography}[{\includegraphics[width=1in,height=1.25in,clip,keepaspectratio]{Xiaofeng_LIU.jpeg}}]{Xiaofeng Liu} (Member, IEEE) is an Assistant Professor at Harvard, Associated Scientist at Broad Institute of MIT and Harvard, and Research staff at Massachusetts General Hospital. His Ph.D. was jointly supervised by advisors from University of Chinese Academy of Science and Carnegie Mellon University. He received the B.Eng. and B.A. from the University of Science and Technology of China in 2014. He was a recipient of the Best Paper award of the IEEE ISBA 2018. He is the program committee or reviewer for PAMI, TIP, TNNLS, NeurIPS, ICML, CVPR, ICCV, ECCV, Nat. Com. His research interests include deep learning, computer vision, and pattern recognition.
% \end{IEEEbiography}

\begin{IEEEbiography}[{\includegraphics[width=1in,height=1.25in,clip,keepaspectratio]{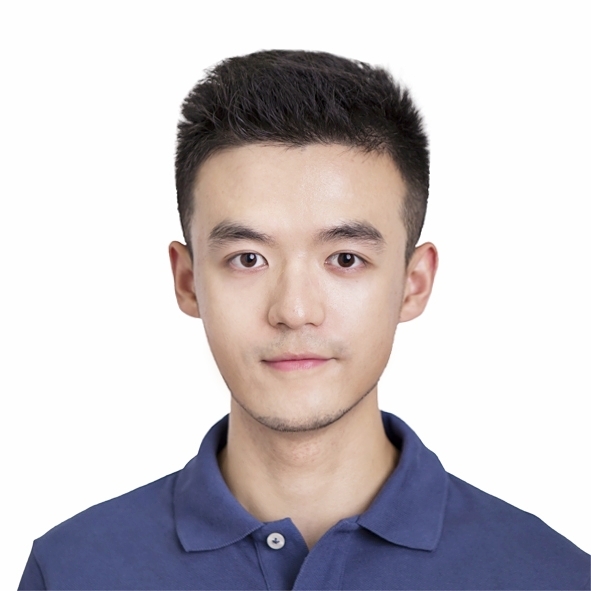}}]{Xiaofeng Liu} (Member, IEEE) is an Assistant Professor at Yale, and an Associate Member at the Broad Institute of MIT and Harvard. He is also an affiliate faculty member of the Center for Biomedical Data Science, Yale Institutes for Foundations of Data Science, and Global Health. His Ph.D. was jointly supervised by advisors from University of Chinese Academy of Science and Carnegie Mellon University. He received the B.Eng. and B.A. from the University of Science and Technology of China in 2014. He was a recipient of the Best Paper award of the IEEE ISBA 2018. He is the program committee or reviewer for PAMI, TIP, TNNLS, NeurIPS, ICML, CVPR, ICCV, ECCV, Nat. Com. His research interests are centered around the convergence of trustworthy AI/deep learning, medical imaging, and data science to advance the diagnosis, prognosis, and treatment monitoring of various diseases.
\end{IEEEbiography}

\begin{IEEEbiography}[{\includegraphics[width=1in,height=1.25in,clip,keepaspectratio]{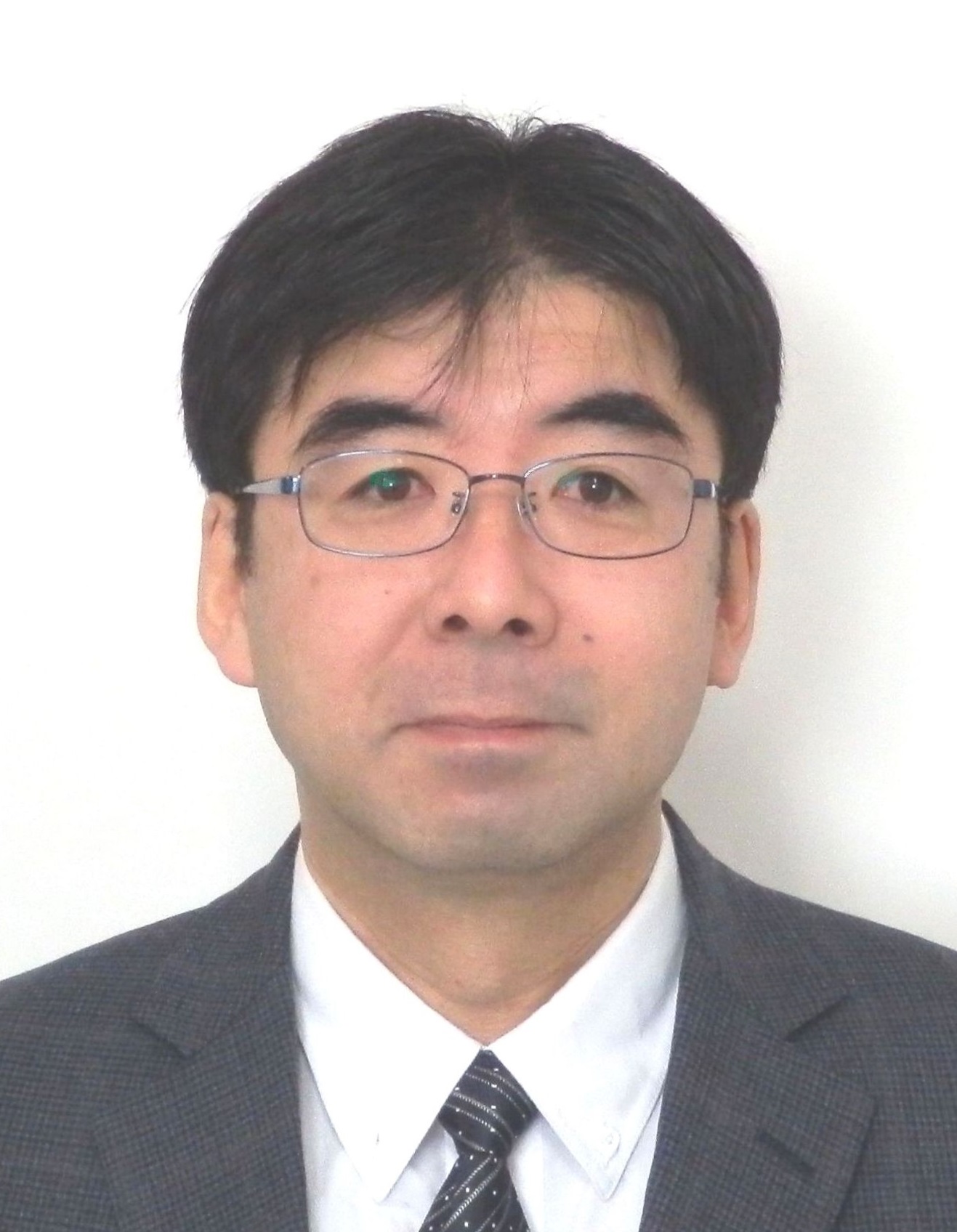}}]{Shinichiro Omachi} 
(Senior Member, IEEE) received the B.E., M.E., and Ph.D. degrees in information engineering from Tohoku University, Japan, in 1988, 1990, and 1993, respectively. He worked as an Assistant Professor at the Education Center for Information Processing,
Tohoku University, from 1993 to 1996. Since 1996, he has been affiliated with the Graduate School of Engineering, Tohoku University, where he is currently a Professor. From 2000 to 2001, he was a Visiting Associate Professor at Brown University.
His research interests include pattern recognition, computer vision, image processing, image coding, and parallel processing. He is a member of the Institute of Electronics, Information and Communication Engineers, the Information Processing Society of Japan, among others. He received the IAPR/ICDAR Best Paper Award in 2007, the Best Paper Method Award of the 33rd Annual Conference of the GfKl in 2010, the ICFHR Best Paper Award in 2010, and the IEICE Best Paper Award in 2012. From 2020 to 2021, he was the Vice Chair of the IEEE Sendai Section. He served as the Editor-in-Chief for IEICE Transactions on Information and Systems from 2013 to 2015.
\end{IEEEbiography}

\newpage

\section*{Appendix}

\hyss{To generate IR images, we first required the surface radiance from the object and then generated the image according to the chosen coloring method. In this section, we will present a classical generation model based on atmospheric physics\cite{yu1998infrared}.}

Atmospheric physics informs us that infrared radiation exhibits varying propagation characteristics at different wavelengths. In many spectral ranges, this radiation experiences significant attenuation, rendering the source radiation detectable only within spectral ranges where the atmosphere is transparent, known as the atmospheric window. Consequently, when simulating infrared images of an object, object surface's radiance can be computed within this atmospheric window, utilizing \textit{Planck's Law}.

The expression $e^{C_1 / \lambda}\gg 1$ holds true within the infrared region, where $\lambda$ is the wavelength: $C_1$ is the second radiation constant, $C_1=1.4388 \times 10^{-2} \mathrm{~m} \cdot \mathrm{K}$. Consequently, we can employ an approximation of \textit{Planck's Law}:

\begin{equation}
E \approx \int_{\lambda_2}^{\lambda_1} \frac{C_1}{\lambda^S} e^{-C_2 / \lambda T} d \lambda
\label{eq4}
\end{equation}

Eq.\ref{eq4} represents the spectral radiant energy emission per unit time and per unit area from a blackbody at wavelength $\lambda$ within the wavelength range $d \lambda$. Here, $C_2$ is the first radiation constant, with a value of $3.742 \times 10^{-16} \mathrm{~W} \cdot \mathrm{m}^2$. $E$ denotes the spectral radiant emittance of a blackbody at an absolute temperature $T$.

By employing the partial integral method, we can approximate the evaluation of this integral as follows:

\vspace{-0.2cm}
\begin{equation}
\scalebox{0.8}{% 缩小比例为 0.8
$E \approx \frac{C_1}{C_2 / T} e^{-C_2 X / T}\left\{X^3+\frac{3}{C_2 / T}\left[X^2+\frac{2}{C_2 / T}\left(X+\frac{1}{C_2 / T}\right)\right]\right\}_{X=\frac{1}{\lambda_1}}^{X=\frac{1}{\lambda_2}}$
}
\label{eq5}
\end{equation}

\hyss{In the field of infrared image synthesis, high-quality IR images rely on a trusted heat transfer model first. This model requires a global estimate on the object's heat transfer state, both inside and on the surface of the object as well as the surrounding environment. Then, we further apply a pending radiance estimation method to calculate the radiance from the object. Finally the coloring model will color the image based on the radiance prior knowledge.}

In accordance with \textit{Wien's Displacement Law}, if $\lambda T<3000 \mu m \cdot K$, the relative error introduced by Eq.\ref{eq5} is less than one percent. Upon determining the radiance of each surface patch using Eq.\ref{eq5}, we compute the radiance of each vertex through interpolation. Ultimately, each surface patch is rendered using Gouraud Shading.

\begin{table}[h]
\centering
\caption{Abbreviations.}
\label{Abbreviations}
\renewcommand\arraystretch{1.5}
\begin{tabular}{cc}
\hline
\multicolumn{1}{c|}{\textbf{Abbreviation}} & \textbf{Full name}                 \\ \hline
\rowcolor[HTML]{EFEFEF} 
SR                                         & super-resolution                   \\
\rowcolor[HTML]{EFEFEF} 
IR                                         & infrared                           \\
\rowcolor[HTML]{EFEFEF} 
HR                                         & high-resolution                    \\
\rowcolor[HTML]{EFEFEF} 
LR                                         & low-resolution                     \\ \hline
JPEG                                       & joint photographic experts group   \\
NUC                                       & non-uniformity correction  \\
IQA                                        & image quality assessment           \\
FLIR                                       & forward-looking infrared           \\
LED                                        & light-emitting diode               \\ \hline
\rowcolor[HTML]{EFEFEF} 
CNN                                        & convolutional neural network       \\
\rowcolor[HTML]{EFEFEF} 
GAN                                        & generative adversarial net         \\
\rowcolor[HTML]{EFEFEF} 
WGAN                                       & wasserstein GAN                    \\
\rowcolor[HTML]{EFEFEF} 
SRFBN                                     & feedback network for image SR                             \\
\rowcolor[HTML]{EFEFEF} 
SRCNN                                      & SR CNN                             \\
\rowcolor[HTML]{EFEFEF} 
ResNet                                     & residual network                           \\
\rowcolor[HTML]{EFEFEF} 
DCGAN                                     & deep convolution GAN                           \\
\rowcolor[HTML]{EFEFEF} 
PSRGAN                                    & progressive SR GAN                           \\
\rowcolor[HTML]{EFEFEF} 
CycleGAN                                     & cycle-in-cycle GAN                           \\
\rowcolor[HTML]{EFEFEF} 
SRGAN                                      & SR GAN                             \\ \hline
EO                                         & electro-optical                    \\
FFT                                         & fast Fourier transform                    \\
FPN                                        & fixed pattern noise                \\
NDTI                                       & normalized difference target index \\
LCTM                                       & land cover target mapping          \\ \hline
\rowcolor[HTML]{EFEFEF} 
PSNR                                       & peak signal-to-noise ratio         \\
\rowcolor[HTML]{EFEFEF} 
SNR                                       & signal-to-noise ratio         \\
\rowcolor[HTML]{EFEFEF} 
SSIM                                       & structural similarity index        \\
\rowcolor[HTML]{EFEFEF} 
MSE                                        & mean square error                  \\
\rowcolor[HTML]{EFEFEF} 
NIQE                                       & natural image quality evaluator    \\
\rowcolor[HTML]{EFEFEF} 
NSS                                        & natural scene statistics           \\
\rowcolor[HTML]{EFEFEF} 
LPIPS                                      & 

perceptual image patch similarity  \\ \hline

\end{tabular}%

\end{table}

\begin{table}[h]
\centering
\caption{Notations.}
\label{Notations}
\renewcommand\arraystretch{1.5}
\resizebox{\columnwidth}{!}{%
\begin{tabular}{cc}
\hline
\multicolumn{1}{c|}{\textbf{Notation}}                                                                & \textbf{Description}                            \\ \hline
\rowcolor[HTML]{EFEFEF} 
$I_{LR},X$                                                                                              & low-resolution image                            \\
$I_{HR}$, $I$                                                                                         & high-resolution image                           \\
\rowcolor[HTML]{EFEFEF} 
$\hat{I},Y$                                                                                             & reconstructed image                             \\
$\mathbb{D}$                                                                                          & degradation function                            \\
\rowcolor[HTML]{EFEFEF} 
$\otimes$                                                                                             & convolution operation                           \\
${k}$                                                                                                 & blur kernel                                     \\
\rowcolor[HTML]{EFEFEF} 
$\downarrow _{d}$                                                                                     & downsampling factor                             \\
$\mathcal{L}$                                                                                         & loss function                                   \\
\rowcolor[HTML]{EFEFEF} 
$\Phi$                                                                                        & regularization term                             \\
$\lambda$                                                                                             & punishment parameter                            \\
\rowcolor[HTML]{EFEFEF} 
$\theta,\Theta$                                                                                       & parameters of model                             \\
$\mathbb{R}$                                                                                          & image patch                                     \\
\rowcolor[HTML]{EFEFEF} 
$\left[\boldsymbol{\Psi}_c^l, \mathbf{\Psi}^l\right],\left[\boldsymbol{\Phi}_c, \mathbf{\Phi}\right]$ & dictionary pairs                                \\
$u,v$,$\mathcal{Z}$                                                                                                 & sparse codes                                    \\
\rowcolor[HTML]{EFEFEF} 
$M,N$                                                                                                 & number of dimensions                            \\
$i$                                                                                                   & $i$ th training image                           \\
\rowcolor[HTML]{EFEFEF} 
$\widetilde{R}$                                                                                       & estimated residual image                        \\
$R$                                                                                                   & true residual image                             \\
\rowcolor[HTML]{EFEFEF} 
$\mathcal{L}_1$                                                                                       & loss functions of forward regression            \\
$\mathcal{L}_2$                                                                                       & loss functions of inverse regression            \\
\rowcolor[HTML]{EFEFEF} 
$\mathcal{L}_{\text {Sobel }}$                                                                        & loss functions of Sobel edge detector           \\
$G$                                                                                                   & generator                                       \\
\rowcolor[HTML]{EFEFEF} 
$D$                                                                                                   & discriminator                                   \\
$I_L$                                                                                                 & generated low-resolution images                 \\
\rowcolor[HTML]{EFEFEF} 
$\mathcal{L}_{\text {WGAN}}$                                                                          & loss functions of WGAN                          \\
$E$                                                                                                   & cross-entropy                                   \\
\rowcolor[HTML]{EFEFEF} 
$p_{data}$                                                                                            & input image data distribution                   \\
$Z$                                                                                                   & distribution of output fake sample by generator \\
\rowcolor[HTML]{EFEFEF} 
$d_{\mathcal{K} \mathcal{L}}$                                                                         & $\mathcal{KL}$ distance  \\
$\nabla$                                                                                              & gradient operator                               \\
\rowcolor[HTML]{EFEFEF} 
$\alpha, \beta, \gamma$                                                                               & control parameters                              \\

$\boldsymbol{\Omega}$,$u$                                                                              & $\ell_1$-norm                            \\ 
\rowcolor[HTML]{EFEFEF} 
$\mathcal{C}_l$                                                                               & brightness similarity                              \\ 
$\mathcal{C}_c$                                                                               & contrast similarity                              \\ 
\rowcolor[HTML]{EFEFEF} 
$\mathcal{C}_s$                                                                               & structural similarity                              \\ 
$P$                                                                               & phase spectrum                              \\
\rowcolor[HTML]{EFEFEF} 
$\lambda_1$ \& $\lambda_2$                                                                               & eigenvalues                              \\ 
$\sigma_{\text {sum}}$ \& $\sigma_{\text {ratio}}$                                                                              & pre-defined parameters                             \\
\rowcolor[HTML]{EFEFEF} 
$\varepsilon$                                                                              & constant                             \\ 
$e$                                                                              & edges                             \\ 
\rowcolor[HTML]{EFEFEF} 
$(\hat{k},l)$                                                                              & pixel                             \\ 
\hline
\end{tabular}%
}
\end{table}

\end{document}